\documentclass[11pt, oneside]{article}   	

\pdfoutput=1
\usepackage{jheppub}

\usepackage{tikz}
\usetikzlibrary{decorations.pathmorphing}

\usepackage[english]{babel}
\usepackage[T1]{fontenc}
\usepackage{lmodern}
\usepackage{physics}
\usepackage{graphicx}
\usepackage{comment}
\usepackage{xspace}
\usepackage{amsmath}				
\usepackage{multirow}				
\usepackage{color}
\usepackage{mathtools}
\usepackage{amssymb}
\usepackage{amsfonts}
\usepackage{hyperref}
\usepackage{esvect}
\usepackage{titlesec}
\usepackage[normalem]{ulem}

\hypersetup{
    bookmarks=true,%
    colorlinks,%
    citecolor=blue,%
    filecolor=black,%
    linkcolor=blue,%
    urlcolor=blue
}

\usepackage{overpic}
\usepackage{array}
\usepackage{rotating}

\usepackage{amssymb}

\title{
New Rotating Black Holes in String Theory}
\author{Watse Sybesma$^*$ and}
\author{Poula Tadros$^{*,\dagger}$}

\affiliation[*]{Nordita, KTH Royal Institute of Technology and Stockholm University,\\ Hannes Alfvéns väg 12, SE-106 91 Stockholm, Sweden}  
\affiliation[\dagger]{Institute of Theoretical Physics, Faculty of Mathematics and Physics, Charles University,\\
V Hole\v{s}ovi\v{c}k\'ach 2, Prague 180 00, Czech Republic}
\abstract{
We present new black hole solutions to the low-energy effective action of string theory. We introduce three- and four-dimensional solutions that are rotating, asymptotically flat, and exhibit a linear dilaton vacuum. 
We also introduce higher-dimensional generalisations of these black holes that possess multi angular momentum-like charges. 
These solutions cannot be overspun, i.e., do not have an extremality condition, akin to some higher-dimensional Myers-Perry black holes. 
Studying their thermodynamics reveals that the temperature associated to these solutions does not depend on the black hole mass, similar to the Witten black hole. 
We also find that their asymptotic symmetry group is more stringent than the BMS group.
We consider the charged generalisations for these black holes, which introduces closed timelike curves within the inner horizon. We show that these black holes can be derived from the large-$d$ limit of the Myers-Perry black hole. As such we advocate that large-$d$ can provide a useful vantage point to interpret the here introduced black holes, as well as more generally a way to generate new effective field theories and corresponding non-trivial solutions.
}

\begin{document}
\maketitle

\section{Introduction}
The study of low-energy string effective field theory (EFT) started as early as the first string revolution where string theory was shown to be a viable quantum gravity model \cite{Green:1984sg}. The necessary condition for string theory to be consistent as a quantum field theory (QFT) is that the two-dimensional worldsheet energy-momentum tensor is traceless; in other words, all beta functions have to vanish. 
For bosonic strings in $26$ dimensions, to the linear order in the Regge slope (the reciprocal of the string's tension), these conditions can be derived from the variation of the action \cite{Callan:1985ia}
\begin{equation}\label{EFT acton}
    S 
    = 
    \frac{1}{16\pi G_{D}}
    \int \text{d}^{D}x \sqrt{g}e^{-2\Phi}\left(R + 4 (\nabla \Phi)^2 - \tfrac{1}{12}H^2\right),
\end{equation}
where $D$ is the number of dimensions on which string theory is defined (in this case $D=26$, and $D=10$ for superstring theory), $g_{\mu\nu}$ is the metric and $R$ the corresponding Ricci scalar, $H_{\mu\nu\sigma} = 3\nabla_{[\mu}B_{\nu\sigma]}$ is the tensor field strength for the antisymmetric tensor $B_{\mu\nu}$, and $\Phi$ is a dilaton field. 
For the purposes of this work we use $G_{D}$ to denote the $D$-dimensional Newton's constant that is related to the string tension and coupling.
The same action was derived as the tree level EFT approximation of bosonic string theory \cite{Fradkin:1984pq}, meaning that this action represents a consistent QFT derived as a low-energy EFT of bosonic string theory in $26$ dimensions. Adding a fermionic sector and requiring supersymmetry, thus defining a superstring model, reduces the critical dimension from $26$ to $10$, but the above action remains the universal common model independent EFT action for all string models to the lowest order.

Since our universe is four-dimensional rather than ten-dimensional, an extensive effort is made to reduce superstring models and their low-energy EFTs to lower dimensions. 
Many compactifications to string theory were proposed \cite{Giveon:1994fu,Fraiman:2018ebo,PhysRevD.35.648,Metsaev:1998it,Hubsch:1986ny,Choi:2004vb,Dixon:1985jw,DIXON1986285,Candelas:1985en} and each choice for an internal manifold results in a different tower of particles in the lower-dimensional model, i.e., different phenomenology. However, the common NS-NS sector is the action \eqref{EFT acton} with $D=4$ or $D=3$ \cite{Gibbons:1994vm}. 
For non-critical string models \cite{Ellis:2005qa}, after integrating out the $B$-field, we arrive at the action
\begin{equation}\label{eq:eff_intro}
	S_{\text{eff}}
	=
	\frac{1}{16\pi G_{D}}
	\int \text{d}^{D}x\sqrt{-g}e^{-2\Phi}
	\left[
		R
		+
		4(\partial\Phi)^{2}
		+
		4\lambda^{2}
	\right]
	\,,
\end{equation}
where $\lambda$ is an energy scale. In this article we study the action above and its solutions in three and four dimensions without further referring to its string theoretical origin -- we will view it as a mere theory of dilaton gravity.
The main novelty we present are rotational black hole solutions with $D=3$ and $D=4$ that, together with a dilaton solution, solve the equations of motion of the action above. We also present charged generalisations. For $D=3$ the metric is given by
\begin{equation}\label{eq:metric}
  \begin{aligned}
    \text{d}s^2 
    =&~ 
    -\text{d}t^2 
    + 
    \left(
    \frac{1+\alpha^2}{1+\alpha^2\cos^2\theta_0} 
     -
     f(w)
     \right)
    (\text{d}t+ \ell \alpha \sin^2\theta_0\text{d}\phi)^2  
 + 
     \frac{\ell^2 \text{d}w^2}{
   	f(w)} 
    +
     \ell^2 (1+\alpha^2) \sin^2\theta_0\text{d}\phi^2
        \,,
\end{aligned}  
\end{equation}
where the emblackening factor $f(w)$ and length scale $\ell$ are given by
\begin{equation}
	f(w)
	=
	\frac{1+\alpha^2}{1+\alpha^2\cos^2\theta_0} 
	\left(
	1
	-
        me^{-w}
        \right)
        \,,
        \quad
        \ell^{2}=\frac{1+\alpha^{2}}{4\lambda^{2}(1+\alpha^{2}\cos^{2}\theta_{0})}
        \,.
\end{equation}
Let us unpack the above solution. 
At the radial location $w=\log(m)$ a coordinate singularity occurs, whereas $w\to-\infty$ signals a true curvature singularity. 
The integration constant $m$ is related to the mass of the black hole, whereas the integration constant $\alpha$ is related to the angular momentum of the black hole. Turning off $\alpha$ yields a metric that is the two-dimensional Witten black hole \cite{Mandal:1991tz,Witten:1991yr} times a circle in the $\phi$ coordinate.
The integration constant $0<\theta_{0}<\pi$ can be interpreted as warping the vacuum asymptotics.
Some intriguing features of this black hole are that its temperature, akin to the Witten black hole, is entirely controlled by the length scale $\ell$ and that similar to Myers-Perry black holes, this black hole can not be overspun; it does not posses an extremal Kerr-like bound. 

In addition to the metric, to fully satisfy the equations of motion the dilaton $\Phi$ also requires a specific solution. It turns out that the solution reveals an asymptotically linear dilaton vacuum:
\begin{equation}\label{eq:linearD3}
\left[4\ell^{2}(\partial\Phi)^{2}\right]_{D=3}
	=
	\frac{1+\alpha^{2}}{1+\alpha^{2}\cos^{2}\theta_{0}}
	\left(
		1
		-
		me^{-w}
	\right)
	\,.
\end{equation}
Linear dilaton spacetimes notably feature in non-AdS/CFT realisations of holography where in particular the $D=2$ version of the low-energy effective action has been explored, see e.g. \cite{Aharony:1998ub,Maldacena:1997cg}.
We also present an uplift of this three-dimensional solution to four dimensions, which requires and extra $\ell^2 (1+\alpha^2 \cos^2\theta_0)\text{d}\theta^2 $ term in the metric and yields a cylindrical black hole solution with the same qualitative features.
We can extend these solutions to an electrically charged black hole by adding a non-minimally coupled Maxwell field to the action. Finally, we also present a solution that possesses multi angular momentum-like charges and is valid for any $D>2$.

The here presented solutions were generated by exploring the large-$d$ and near-horizon limit of Myers-Perry black holes in combination with an S-wave reduction of the transverse spacetime.\footnote{Complementary large-$d$ aspects of Myers-Perry black holes were studied in \cite{Emparan:2014jca,Caldarelli:2010xz,Andrade:2018nsz,Andrade:2018rcx}. See \cite{Emparan:2020inr} for a comprehensive review on the large-$d$ programme. }
Previously, in \cite{Soda:1993xc,Emparan:2013xia} the emergence of the low-energy effective action of string theory in two dimensions was noticed by considering a large-$d$ limit of the Schwarzschild-Tangherlini black hole. In \cite{Sybesma:2022nby} this method was extended by considering the large-$d$ limit of a Lifshitz black hole, which yielded a family of two-dimensional models that interpolates between the model found by \cite{Emparan:2013xia} and Jackiw-Teitelboim gravity \cite{JACKIW1985343,TEITELBOIM198341} and was further studied in \cite{Harksen:2024uik} for its holographic properties.
On the one hand the large-$d$ perspective offers a vantage point from which to interpret the lower-$d$ quantities. For example, the parameters $m$ and $\alpha$ are related to their higher dimensional counterparts, and less obviously, the parameter $\theta_{0}$ arises from zooming in around a polar value $\theta=\theta_{0}$.
On the other hand, we advocate that the large-$d$ perspective can be used as method of generating new effective models and solutions. While the solutions to the models found in \cite{Emparan:2013xia} and \cite{Sybesma:2022nby} were known in the literature, the solutions we discuss here are novel.

We present our work in the following order. We start by explicitly showing how the here presented $D=3$ and $D=4$ solutions solve the equations of motion of the low-energy effective action in Section \ref{sec:2}. In Section \ref{sec:3} we study properties of these solutions, such as their thermodynamics. 
We elaborate on the large-$d$ pedigree in Section \ref{sec:4}.
In Section \ref{newsec5} we present a black hole solution that possesses multi angular momentum-like charges and discuss some of its properties. 
A discussion of the results and ideas for future work are presented in Section \ref{sec:5}.
\section{Black Hole Solutions}\label{sec:2}
We are interested in solving the equations of motion of the action \eqref{eq:eff_intro} for $D=4$ and $D=3$, which will be the main object of study in this article due to its simplicity. We will also consider the possibility for the black hole solution to carry charge, by introducing a Maxwell field to the action. We start with presenting the four-dimensional case in Section \ref{sec:4daction} and then turn to Section \ref{sec:3dmodel} for the three-dimensional case.
%
\subsection{The four-dimensional solution}\label{sec:4daction}
%
Let us consider the four-dimensional version of the low-energy effective string action in \eqref{eq:eff_intro}: 
\begin{equation}
    S_{4d,Q} 
    = 
    \frac{1}{16\pi G_{4}}
    \int \text{d}^4 x \sqrt{-g}~e^{-2\Phi} \left[ R + 4 (\partial \Phi)^2 +4\lambda^{2} - \frac{1}{4} F^2\right],
\end{equation}
where we added a Maxwell field to be able to account for charged solutions: $F^2 = F_{\mu\nu}F^{\mu\nu}$ and $F_{\mu\nu}=\nabla_{\mu}A_{\nu}-\nabla_{\nu}A_{\mu}$.
Let us now present the relevant equations of motion.
From this action, we can derive the equations for the Maxwell field:
\begin{equation}\label{eq:maxwell1}
    	\nabla_{\mu}\left(e^{-2\Phi}F^{\mu\nu}\right)
	=
	0
	\,.
\end{equation}
The Einstein's equations read:
\begin{equation}\label{eq:einstein1}
    R_{\mu\nu} 
    - 
    \frac{1}{2}g_{\mu\nu}R 
    + 
    2 g_{\mu\nu}(\partial \Phi)^2 
    + 
    2\nabla_{\mu}\nabla_{\nu}\Phi 
    - 
    2g_{\mu\nu} \Box \Phi 
    - 
    2\lambda^{2}g_{\mu\nu} 
    = 
    \frac{1}{2} F^{\sigma}{_\mu}F_{\sigma\nu} 
    - 
    \frac{1}{8}g_{\mu\nu}F^2
    \,.
\end{equation}
The dilaton equation of motion becomes:
\begin{equation}\label{eq:dilaton4d}
    R 
    - 
    4(\partial \Phi)^2 
    + 
   4\lambda^{2}
    + 
    4 \Box \Phi 
    - 
    \frac{1}{4}F^2 = 0
    \,.
\end{equation}
Our claim is that the equations of motion above are solved by the following:
\begin{equation}\label{eq:chargedmetric}
  \begin{aligned}
    \text{d}s^2 
    =&~ 
    -\text{d}t^2 
    + 
    \frac{1+\alpha^2}{1+\alpha^2\cos^2\theta_0} 
    \left( 
        me^{-w}
        -
        Q^2e^{-2w}
    \right)
    (\text{d}t+ \ell \alpha \sin^2\theta_0\text{d}\phi)^2  
    \\&~ + 
    \frac{1+\alpha^2\cos^2\theta_0}{1+\alpha^2} \frac{\ell^2 \text{d}w^2}{
   	1
	-
	me^{-w}
        +
        Q^2e^{-2w}} 
    +
     \ell^2 (1+\alpha^2) \sin^2\theta_0\text{d}\phi^2
    \\&~
    + 
    \ell^2 (1+\alpha^2 \cos^2\theta_0)\text{d}\theta^2 
        \,,
	\\
    A
    =&~
    Qe^{-w} \frac{ \sqrt{2(1+\alpha^{2})}}{\sqrt{1+\alpha^{2}\cos^{2}\theta_{0}}}
    \left(
    	\text{d}t
	+
	\ell\alpha \sin^{2}\theta_{0}
	\text{d}\phi
    \right)\,,
	\\
    e^{-2\Phi}
    =&~
    e^{w-\theta \tan\theta_{0}}
    \,,
    \end{aligned}
\end{equation}
with (in the case of four dimensions)
\begin{equation}
	\ell^{2}= \frac{1}{4\lambda^{2}\cos^{2}\theta_{0}}
	\,.
\end{equation}
We leave analysis of the interpretation of the parameters $m$, $Q$, and $\alpha$ for the next section, where we present a further analysis of this geometry. 

Using Mathematica we checked that the equations of motion are solved using the claimed solutions. For transparency and reproducibility we show some intermediate results -- although we stress we will derive these solutions in Section \ref{sec:4} from a large-$d$ perspective.
Let us consider the Maxwell equations \eqref{eq:maxwell1}. Computing
\begin{equation}
	F^{\mu\nu}
	=
	\frac{4Qe^{-w}}{\ell^{2}}
	\left(
	\frac{1+\alpha^{2}}{
		2+\alpha^{2}+\alpha^{2}\cos 2\theta^{0}
	}
	\right)^{\frac{3}{2}}
	\begin{pmatrix}
		0&1&0&0\\
		-1&0&0&\frac{\alpha}{1+\alpha^{2}}\frac{1}{\ell}\\
		0&0&0&0\\
		0&-\frac{\alpha}{1+\alpha^{2}}\frac{1}{\ell}&0&0
	\end{pmatrix}
	\,,
\end{equation}
we see that the only explicit coordinate dependence in $e^{-2\Phi}F^{\mu\nu}$ is $\theta$. Due to the anti-symmetric structure of $F^{\mu\nu}$, we see that $\partial_{\mu}\left(e^{-2\Phi}F^{\mu\nu}\right)=0$ and conclude the Maxwell equations are satisfied.

To solve the Einstein's equations \eqref{eq:einstein1} and the dilaton equation of motion in \eqref{eq:dilaton4d}, the following expressions are helpful
\begin{equation}
	R_{\mu\nu}
	-
	\frac{1}{2}R
	g_{\mu\nu}
	=
	\begin{pmatrix}
		-\frac{\alpha^{2}\sin^{2}\theta_{0}}{2+\alpha^{2}+\alpha^{2}\cos2\theta_{0}}&0&0&-\frac{\alpha^{2}\sin^{2}\theta_{0}}{2+\alpha^{2}+\alpha^{2}\cos2\theta_{0}}\\
		0&0&0&0\\
		0&0&-\frac{\ell^{2}}{4}\left(2+\alpha^{2}+\alpha^{2}\cos2\theta_{0}\right)&0\\
		-\frac{\alpha^{2}\sin^{2}\theta_{0}}{2+\alpha^{2}+\alpha^{2}\cos2\theta_{0}}&0&0&-\frac{\alpha^{2}\sin^{2}\theta_{0}}{2+\alpha^{2}+\alpha^{2}\cos2\theta_{0}}
	\end{pmatrix}
	R
	\,,
\end{equation}
with the Ricci scalar given by
\begin{equation}
	R
	=
	\frac{1}{\ell^{2}}
	\frac{2(1+\alpha^{2})}{2+\alpha^{2}+\alpha^{2}\cos^{2}2\theta_{0}}
	\left(
		e^{-w}m
		-
		4Q^{2}e^{-2w}
	\right)
	\,,
\end{equation}
and
\begin{equation}
	\Box
	\Phi
	=
	\frac{e^{-2w}}{\ell^{2}}
	\left(
		2Q^{2}
		-
		me^{w}
	\right)
	\frac{1+\alpha^{2}}{2+\alpha^{2}+\alpha^{2}\cos2\theta_{0}}
	\,,
\end{equation}
\begin{equation}
	g_{ww}\nabla_{\mu}\nabla_{\nu}\Phi
	=
	\begin{pmatrix}
		-\frac{\ell^{2}}{2}&0&0&-\frac{\ell^{3}}{2}\alpha \sin^{2}2\theta_{0}\\
		0&\frac{e^{4w}\ell^{4}(2+\alpha^{2}+\alpha^{2}\cos2\theta_{0})^{2}}{8(e^{2w}-e^w m+Q^2)^{2}(1+\alpha^{2})^{2}}&0&0\\
		0&0&0&0\\
		-\frac{\ell^{3}}{2}\alpha \sin^{2}2\theta_{0}&0&0&-\frac{\ell^{4}}{2}\alpha^{2} \sin^{4}\theta_{0}
	\end{pmatrix}
	\Box
	\Phi
	\,,
\end{equation}
as well as
\begin{equation}
	4\ell^{2}(\partial\Phi)^{2}
	=
	\frac{1+\alpha^{2}}{1+\alpha^{2}\cos^{2}\theta_{0}}
	\left(
		1
		-
		me^{-w}
		+
		Q^{2}e^{-2w}
	\right)
	+
	\frac{
		\tan^{2}\theta_{0}
	}{	
		1+\alpha^{2}\cos^{2}\theta_{0}
	}
	\,.
\end{equation}
This last line also highlights the fact that this solution indeed possesses a linear dilaton vacuum.

Finally, we can see that by taking the limit $w \to \infty$ (or in fact turning off the parameters $Q$ and $m$), the resulting metric is
\begin{equation}
    \text{d}s^2 = 
    -\text{d}t^2 
    + 
    \frac{1+\alpha^2 \cos^2\theta_0}{1+\alpha^2}\ell^2 \text{d}w^2 
    + 
    \ell^2 (1+\alpha^2)\sin^2\theta_0 \text{d}\phi^2 
    + 
    \ell^2 (1+\alpha^2 \cos^2\theta_0)\text{d}\theta^2,
\end{equation}
which, after rescaling $t,w,\theta$ to some parameters $\tilde{t},\tilde{w},\tilde{\theta}$ can be written as
\begin{equation}\label{w to infty metric}
    \text{d}s^2 = \ell^2 (1+\alpha^2)\sin^2\theta_0\left(-\text{d}\tilde{t}^2 + \text{d}\tilde{w}^2+\text{d}\phi^2+\text{d}\tilde{\theta}^2 \right).
\end{equation}
This last metric is a warped Minkowski$_3$ $\times$ $S^1$. It is for this reason that we can interpret the parameter $\theta_0$ as related to the size of the space as it appears in the warping factor, in addition to the large-$d$ perspective where it signifies a $\hat{\theta}$ value (the higher $d$ polar coordinate) around which is zoomed in as large $d$ is taken, as we will see. A similar interpretation holds for the three-dimensional geometry we analyse next. 

%
\subsection{The three-dimensional solution}\label{sec:3dmodel}
%
We will now consider the three-dimensional version of the low-energy effective string action in \eqref{eq:eff_intro}.
Starting from the four-dimensional action in the previous section, we now aim to integrate out $\theta$ in order to obtain a three-dimensional solution. This requires us to use the on-shell solution
\begin{equation}
	\Phi(w,\theta)
	=
	\Phi(w)
	+
	\frac{\tan\theta_{0}}{2}\theta
	\,,
\end{equation}
and the metric from \eqref{eq:chargedmetric}:
\begin{equation}
    \text{d}s^2 
    =
  	g_{\mu\nu}
    \text{d}x^{\mu}\text{d}x^{\nu}    + 
    \ell^2 (1+\alpha^2 \cos^2\theta_0)\text{d}\theta^2 
        \,,
\end{equation}
where $\mu$ and $\nu$ run over $w$, $t$ and $\phi$.
At the level of the action this translates to effectively keeping the same functional form of the action (assuming the field strength having no legs in the $\theta$ direction)
\begin{equation}\label{eq:3dcharged}
	S_{3d,Q}
	= 
	\frac{1}{16\pi G_{3}}\int \text{d}^3 x \sqrt{-g}~e^{-2\Phi} 
	\left[ 
		R 
		+ 
		4 (\partial \Phi)^2 
		+
		4\lambda^{2} 
		- 
		\frac{1}{4} F^2
	\right]
	\,,
\end{equation}
where $G_{3}^{-1}:=G_{4}^{-1}\int \text{d}\theta e^{-\theta\tan\theta_{0}}\ell\sqrt{1+\alpha^{2}\cos^{2} \theta}$ and (in the case of three dimensions)
\begin{equation}\begin{aligned}
	4\lambda^{2}
	:=&~
	\frac{1}{\ell^{2}\cos^{2}\theta_{0}}
	-
	\frac{\tan^{2}\theta_{0}}{\ell^{2}(1+\alpha^{2}\cos^{2}\theta_{0})}
	\\=&~
	\frac{1+\alpha^{2}}{\ell^{2}(1+\alpha^{2}\cos^{2}\theta_{0})}
	\,.
\end{aligned}\end{equation}
The negative contribution to $4\lambda^{2}$ comes from integrating out the kinetic term in favour of a potential term, which involves carefully flipping a sign in order to retain the same contribution to the equations of motion.
The equations of motion of the three-dimensional action in \eqref{eq:3dcharged} read:
\begin{equation}
    	\nabla_{\mu}\left(e^{-2\Phi}F^{\mu\nu}\right)
	=
	0
	\,,
	\quad
    R 
    - 
    4(\partial \Phi)^2 
    + 
    4\lambda^{2}
    + 
    4 \Box \Phi 
    - 
    \frac{1}{4}F^2 = 0
    \,,
\end{equation}
\begin{equation}
    R_{\mu\nu} 
    - 
    \frac{1}{2}g_{\mu\nu}R 
    + 
    2 g_{\mu\nu}(\partial \Phi)^2 
    + 
    2\nabla_{\mu}\nabla_{\nu}\Phi 
    - 
    2g_{\mu\nu} \Box \Phi 
    - 
    2\lambda^{2}g_{\mu\nu} 
    = 
    \frac{1}{2} F^{\sigma}{_\mu}F_{\sigma\nu} 
    - 
    \frac{1}{4}g_{\mu\nu}F^2
    \,.
\end{equation}
Owing to the dimensional reduction, we know that these solutions are solved by the following three-dimensional solutions: 
\begin{equation}\label{eq:chargedmetric3d}
  \begin{aligned}
    \text{d}s^2 
    =&~ 
    -\text{d}t^2 
    + 
    \frac{1+\alpha^2}{1+\alpha^2\cos^2\theta_0} 
    \left( 
        me^{-w}
        -
        Q^2e^{-2w}
    \right)
    (\text{d}t+ \ell \alpha \sin^2\theta_0\text{d}\phi)^2  
    \\& + 
    \frac{1+\alpha^2\cos^2\theta_0}{1+\alpha^2} \frac{\ell^2 \text{d}w^2}{
   	1
	-
	me^{-w}
        +
        Q^2e^{-2w}} 
    +
     \ell^2 (1+\alpha^2) \sin^2\theta_0\text{d}\phi^2
        \,,
        \\
	\Phi
	=&~
	-\frac{1}{2}w
	\,.
\end{aligned}  
\end{equation}
In \eqref{eq:linearD3} we showed this solution contains a linear dilaton vacuum.
The limit $w\to \infty$ (turning off the black hole) for the three-dimensional black hole is derived directly from \eqref{w to infty metric} by dropping the $\text{d}\tilde{\theta}$ term leaving us with warped Minkowski$_2$ $\times$ $S^1$.

\section{Properties of the Solutions}\label{sec:3}
In this section we study properties of the solutions established in the previous section. In particular, all qualitative results will coincide for the three- and four-dimensional models. Four-dimensional thermodynamical densities (with respect to the $\theta$-direction) can be interpreted as non-density quantities in three dimensions. For this reason, we will without any loss of generality consider the three-dimensional geometry explicitly and only comment on the four-dimensional geometry in order to avoid repetition.

We study important surfaces of the reduced geometries in Section \ref{sec:surfaces}. We then study charges and thermodynamics in Section \ref{sec:charges}. We analyse the symmetries of the metric Section \ref{sec:properties3d}. Finally, we comment on the Penrose process in Section \ref{sec:penrose}.
\subsection{Important surfaces}\label{sec:surfaces}
The three-dimensional charged metric \eqref{eq:3dcharged} has a curvature singularity at $w = -\infty$ where the Ricci scalar diverges: 
\begin{equation}
	R
	=
	\frac{1}{\ell^{2}}
	\frac{1+\alpha^{2}}{1+\alpha^{2}\cos^{2}\theta_{0}}
	\left(
		e^{-w}m
		-
		4Q^{2}e^{-2w}
	\right)
	\,.
\end{equation}
In addition, the Ricci scalar tends to zero as $w\to\infty$.
The geometry possesses two coordinate singularities (horizons) whose locations are obtained by solving $1-m e^{-w}+Q^2e^{-2w}=0$:
\begin{equation}\label{eq:wplus}
	e^{w_{\pm}} 
	= 
	\frac{m \pm \sqrt{m^2-4Q^2}}{2}
	\,.   
\end{equation}
For this black hole there is an extremality condition: 
\begin{equation}
	m^2=4Q^2
	\quad
	\Leftrightarrow
	\quad 
	w_{+}=w_{-}
	\,.
\end{equation} 
Although the black hole is rotating (as we will show in the next subsection), this condition resembles Reissner-Nordstr\"{o}m's extremality condition rather than the Kerr-Newman counterpart pointing out the fact that the extremality condition stems from the added charge and not from rotation (upon setting $Q=0$ we do not have an extremality condition anymore). 
The property of not having an extremality condition from rotation, can be viewed as inherited from the higher dimensional Myers-Perry metric where extremality condition does not exist for $d \geq 6$, and the black hole can not be overspun. This connection we will make more concrete in Section \ref{sec:4}, where we adopt the large-$d$ perspective.
 
 Like any rotating black hole, the geometry has two ergospheres ($E_{\pm}$) whose locations $w_{E_{\pm}}$ can be inferred by solving $g_{tt}(w_{E_{\pm}})=0$ where timelike vectors turn into spacelike when crossing the surface. Thus, no observer can be stationary, with respect to an outside observer, inside the ergosphere. 
 Let us compute the location of the outer (inner) ergo region denoted by $w_{E_{+}}$ ($w_{E_{-}}$). Solving $g_{tt}(w_{E_{\pm}})=0$, one gets
\begin{equation}
	e^{w_{E_{\pm}}} 
	= 
	\frac{1+\alpha^2}{1+\alpha^2\cos^2\theta_0}
	\frac{m \pm \sqrt{m^2 - 4Q^2 \frac{1+\alpha^2\cos^2\theta_0}{1+\alpha^2}}}{2}
    \,.
\end{equation}
It holds that $e^{w_{E_{+}}} > e^{w_+} \geq e^{w_-}$, however, but it is not necessary that $e^{w_-}>e^{w_E{}_{-}}$ (depending on the values of $\alpha$ and $\theta_0$). The existence of an ergosphere outside the outer horizon allows us to extract energy from the black hole using the Penrose process \cite{Penrose:1971uk} as in the case of a Myers-Perry black hole. However, unlike the Myers-Perry black hole, we observe that the inner ergosphere may be outside the inner horizon. 
Though odd, we believe this has no further consequences.

%
We now turn to the study of closed timelike curves that turn out to exist only when we charge the black hole. 
Since the four-dimensional black hole has a line singularity at $w = -\infty$, we expect to possibly have closed timelike curves if we go around this singularity (along a closed $\phi$ curve), which would also persist for the three-dimensional black hole since the $\theta$ direction differentiating between them is trivial. To show this, we set $t,\theta,w$ to constants and let $w=w_{c}$, which we want to yield surfaces separating regions with and without closed timelike curves. We now want to test if $\phi$ curves can be timelike, $\text{d}s^{2}<0$. We start with
 \begin{equation}
 \begin{aligned}
 	\text{d}s^2 
	&=
    \left[
        \frac{1+\alpha^2}{1+\alpha^2\cos^2\theta_0}(m e^{-w_c}
        -Q^2e^{-2w_c})\ell^2 \alpha^2 \sin^4\theta_0+ \ell^2(1+\alpha^2)\sin^2\theta_0
    \right]
    \text{d}\phi^2\\&
     = \ell^2 (1+\alpha^2)\sin^2\theta_0\frac{1+\alpha^2}{1+\alpha^2 \cos^2\theta_0}\left[(m e^{-w_c} - Q^2 e^{-2w_c}) \alpha^2 \sin^2\theta_0 + 1+\alpha^2\cos^2\theta_0 \right],
    \end{aligned}
 \end{equation}
 which can be simplified into (combined with the timelike condition)
 \begin{equation}
     Q^2 e^{-2w_c} - me^{-w_c} >  \frac{1+\alpha^2 \cos^2\theta_0}{\alpha^2\sin^2\theta_0}>0
     \,.
 \end{equation}
To compare $w_c$ to $w_{\pm}$, we note that at $w=w_{\pm}$, $1-me^{-w_{\pm}}+Q^2e^{-2w_{\pm}}=0$, i.e., outside the outer horizon, we have $Q^2 e^{-2w} - me^{-w} \in [-1,0]$, thus, closed timelike curves cannot exist in this region. In the region between the outer and inner horizons, we have $Q^2 e^{-2w} - me^{-w} \leq -1$ so closed timelike curves can not exist there either. However, inside the inner horizon (in the vicinity of the singularity), we have $Q^2 e^{-2w} - me^{-w} \in [-1,\infty)$, meaning that closed timelike curves can occur. Indeed, for $Q=0$, the uncharged case, closed timelike curves are absent, akin to, for example, the Kerr black hole. 
\subsection{Charges and thermodynamics}\label{sec:charges}
Since the four-dimensional metric has a trivial direction (namely, the $\theta$ direction), we will focus only on the three-dimensional geometry without any loss of generality. The asymptotic charges associated with the mass and the angular momentum for the geometry in question are calculated via Komar integrals \cite{Nakarachinda:2022gsb,Modak:2010fn,Banerjee:2010ye}
\begin{equation}
    \begin{aligned}
        M &= -\frac{1}{8\pi G_3} \int_{\substack{w\to \infty\\ t=\text{constant}}}e^{-2\Phi}\star \text{d}\sigma^{(t)},\\
        J &= - \frac{1}{16\pi G_3}\int_{\substack{w\to \infty\\ t=\text{constant}}} e^{-2\Phi}\star \text{d}\sigma^{(\phi)},
    \end{aligned}
\end{equation}
where $\star$ is the Hodge dual operator, $\sigma^{(t)} = g_{tt}\text{d}t + g_{t\phi}\text{d}\phi$ is the dual form of the timelike Killing vector $(\partial_t)^{\mu}$, and $\sigma^{(\phi)} = g_{t\phi}\text{d}t + g_{\phi\phi}\text{d}\phi$ is the dual form of the spacelike Killing vector $(\partial_{\phi})^{\mu}$. Notice that the normalisation constants are different in the mass and angular momentum formulas, which is done to match the Komar mass and angular momentum with the mass and angular momentum in the non-relativistic (Newtonian) limit.  Substituting with the three-dimensional charged metric \eqref{eq:chargedmetric3d}, we get the Komar mass
\begin{equation}\label{eq:mass1}
    \begin{aligned}
    M &= - \frac{1}{8\pi G_3 \ell}\int_{\substack{w\to\infty \\ t=\text{constant}}}e^{w}\frac{\partial g_{tt}}{\partial  w}\sqrt{g_{\phi\phi}}\text{d}\phi\\&
    = \frac{(e^{w_+}+e^{w_-}) |\sin\theta_0|}{4G_3}\frac{(1+\alpha^2)^{3/2}}{1+\alpha^2\cos^2\theta_0}
    \,,
    \end{aligned}
\end{equation}
and the Komar angular momentum
\begin{equation}\label{ang.mom.}
    \begin{aligned}
        J &= -\frac{1}{16\pi G_3 \ell}\int_{\substack{w\to\infty \\ t=\text{constant}}}e^{w}\frac{\partial g_{t\phi}}{\partial w}\sqrt{g_{\phi\phi}}\text{d}\phi\\&
        = \frac{(e^{w_+}+e^{w_-})|\sin \theta_0|}{8G_3}\frac{(1+\alpha^2)^{3/2}}{1+\alpha^2\cos^2\theta_0} \ell \alpha \sin^2\theta_0
        \,.
    \end{aligned}
\end{equation}
Notice that $J$ is proportional to $M$ similar to the relation between mass and angular momentum of the Myers-Perry black hole as reviewed in \eqref{eq:rotandmass} in Section \ref{sec:MP1}:
\begin{equation}
    J = \frac{1}{2} M \ell \alpha \sin^2\theta_0.
\end{equation}

The angular velocity at the horizon is given by the usual formula $\Omega_H = \frac{g_{t\phi}}{g_{\phi\phi}}$ evaluated at the outer horizon. For our metric, the result is
\begin{equation}
    \Omega_H = \frac{\alpha}{\ell (1+\alpha^2)}.
\end{equation}
The entropy of the black hole can be evaluated using Wald's formula \cite{Wald:1993nt}
\begin{equation}\label{entropy}
\begin{aligned}
    S =&~ \frac{1}{4G_3}\int_H e^w \sqrt{g_{\phi\phi}}\text{d}\phi 
    \\=&~ \frac{\pi \ell |\sin\theta_0| e^{w_+}}{2G_3}\frac{1+\alpha^2}{\sqrt{1+\alpha^2\cos^2\theta_0}}
    \,,
\end{aligned}\end{equation}
and the Hawking temperature is given by $T = \frac{\kappa}{2\pi}$, where $\kappa$ is the surface gravity at the horizon, and is calculated via its definition
\begin{equation}
    \xi^{\mu}\kappa = \xi^{\nu}\nabla_{\nu}\xi^{\mu},
\end{equation}
where $\xi^{\mu}$ is the normal vector to the outer horizon, in our case, $\xi^{\mu} = (\partial_t)^{\mu} - \Omega_H (\partial_{\phi})^{\mu}$. 
This yields
\begin{equation}\label{eq:hawkingtemperature}
    T = \frac{1}{4\pi \ell}\left(1-\frac{e^{w_-}}{e^{w_+}}\right).
\end{equation}
Notice that in the uncharged case, i.e., when $e^{w_-}=0$, the temperature is a constant which is a feature it shares with the two-dimensional black hole of string theory, or Witten black hole, which was first studied by Mandal, Sengupta and Wadia \cite{Mandal:1991tz}, and Witten \cite{Witten:1991yr}. An alternative derivation for the temperature and angular velocity can be found in Appendix \ref{appendixA}, which yields the same answers.

Since the black hole in question is charged, we can calculate the total electric charge using Gauss's law
\begin{equation}
    \begin{aligned}
        Q_{\text{total}} &= \frac{1}{4\pi} \int_{\substack{w\to\infty \\ t=\text{constant}}} e^w F^{wt}\sqrt{g_{\phi\phi}}\text{d}\phi
        \,.
    \end{aligned}
\end{equation}
where $F_{\mu\nu}$ is the electromagnetic tensor derived from the vector potential in the solution \eqref{eq:chargedmetric}. Evaluating the integral, we get the total charge
\begin{equation}\label{Qtotal}
    Q_{\text{total}} = \frac{e^{(w_++w_-)/2} |\sin\theta_0|}{\sqrt{2}\ell}\frac{(1+\alpha^2)^2}{(1+\alpha^2\cos^2\theta_0)^{3/2}}.
\end{equation}
The electric potential at the horizon is computed by calculating the work done to move a unit charge from infinity to the outer horizon, i.e.,
\begin{equation}
    \Phi_{\text{el}} = \xi^{\mu}A_{\mu}|_{\infty} - \xi^{\mu}A_{\mu}|_{H} = -\sqrt{2}e^{(w_--w_+)/2}\sqrt{\frac{1+\alpha^2\cos^2\theta_0}{1+\alpha^2}}.
\end{equation}

Having calculated the asymptotic charges and the basic thermodynamical quantities, we now study the thermodynamical stability of the black hole at hand. First of all, the free energy (canonical ensemble) is given by
\begin{equation}
    F(T,J,Q_{\text{total}}) = M-ST,
\end{equation}
where $T$, $J$, and $Q_{\text{total}}$ are kept fixed implicitly through $M$ and $S$, see e.g. \cite{Caldarelli:1999xj}. 
Substituting and doing some algebra, we get
\begin{equation}
\begin{aligned}
 F 
 &= \frac{|\sin\theta_0|}{4G_3}\frac{(1+\alpha^2)^{3/2}}{1+\alpha^2\cos^2\theta_0}\left( e^{w_+}+e^{w_-} -\sqrt{\frac{1+\alpha^2\cos^2\theta_0}{1+\alpha^2}}\frac{e^{w_+}-e^{w_-}}{2}\right)>0.
 \end{aligned}
\end{equation}
Since $e^{w_+}-e^{w_-} \leq e^{w_+}+e^{w_-}$ with the equality being satisfied only in the uncharged case ($e^{w_-}=0$), and $\frac{1}{2}\sqrt{\frac{1+\alpha^2\cos^2\theta_0}{1+\alpha^2}} < 1$, therefore, the free energy is always positive. Thus, the black hole is globally unfavoured.

To test local stability, we calculate the specific heat of the black hole in the canonical ensemble. First of all, we notice that the explicit thermodynamical parameters to be used are the temperature $T$, the angular momentum $J$, and the total electric charge $Q_{\text{total}}$ where $\alpha$, $e^{w_{\pm}}$ are functions of those parameters. Rather than explicitly inverting $\alpha (T,J,Q_{\text{total}})$ and $w_{\pm}(T,J,Q_{\text{total}})$, we only invert $e^{w_-}$ and derive the $T$ derivative of the rest by imposing $J=\text{constant}$ and $Q_{\text{total}}=\text{constant}$, or rather $\text{d}J=\text{d}Q_{\text{total}}=0$.\\
From \eqref{eq:hawkingtemperature}, we have $e^{w_-} = e^{w_+}(1-4\pi \ell T)$. Substituting into 
\eqref{ang.mom.} and \eqref{Qtotal} we get
\begin{equation}\label{thermodynamical quantities}
    \begin{aligned}
        J &= \frac{e^{w_+}(2-4\pi\ell T)|\sin\theta_0|}{8G_3}\frac{(1+\alpha^2)^{3/2}}{1+\alpha\cos^2\theta_0}\alpha\ell\sin^2\theta_0,\\
        Q_{\text{total}} &= \frac{e^{w_+}\sqrt{1-4\pi\ell T}|\sin\theta_0|}{\sqrt{2}\ell}\frac{(1+\alpha^2)^{2}}{(1+\alpha\cos^2\theta_0)^{3/2}}.
    \end{aligned}
\end{equation}
Noting that $J$ and $Q_{\text{total}}$ depend on three variables explicitly ($\alpha$, $T$ and $e^{w_+}$), imposing $\text{d}J = \text{d}Q_{\text{total}}=0$ leads to the two equations 
\begin{equation}
\begin{aligned}
   0=&~ \frac{\partial J}{\partial e^{w_+}}\text{d}e^{w_+} + \frac{\partial J}{\partial \alpha}\text{d}\alpha + \frac{\partial J}{\partial T}\text{d}T  \,,\\
    0=&~\frac{\partial Q_{\text{total}}}{\partial e^{w_+}}\text{d}e^{w_+} + \frac{\partial Q_{\text{total}}}{\partial \alpha}\text{d}\alpha + \frac{\partial Q_{\text{total}}}{\partial T}\text{d}T
    \,.
    \end{aligned}
\end{equation}
From these equations, we get\footnote{Multiplying the first equation by $\frac{\partial Q_{\text{total}}}{\partial \alpha}$ and the second equation by $\frac{\partial J}{\partial \alpha}$, we get
\begin{equation}
    \left(\frac{\partial J}{\partial e^{w_+}}\frac{\partial Q_{\text{total}}}{\partial \alpha} - \frac{\partial J}{\partial \alpha}\frac{\partial Q_{\text{total}}}{\partial e^{w_+}} \right) \text{d}e^{w_+} + \left(\frac{\partial J}{\partial T}\frac{\partial Q_{\text{total}}}{\partial \alpha} - \frac{\partial J}{\partial \alpha}\frac{\partial Q_{\text{total}}}{\partial T} \right)\text{d}T=0\,.
\end{equation}
Then multiplying the first equation by $\frac{\partial Q_{\text{total}}}{\partial e^{w_+}}$ and the second equation by $\frac{\partial J}{\partial e^{w_+}}$, and subtracting the previous equation, we get
\begin{equation}
    \left(\frac{\partial J}{\partial \alpha}\frac{\partial Q_{\text{total}}}{\partial e^{w_+}}-\frac{Q_{\text{total}}}{\partial \alpha}\frac{\partial J}{\partial e^{w_+}} \right)\text{d}\alpha + \left( \frac{\partial J}{\partial T}\frac{\partial Q_{\text{total}}}{\partial e^{w_+}}-\frac{Q_{\text{total}}}{\partial T}\frac{\partial J}{\partial e^{w_+}} \right)\text{d}T=0\,.
\end{equation}}
\begin{equation}
    \begin{aligned}
 \frac{\text{d}e^{w_+}}{\text{d}T}=&~\frac{
\frac{\partial Q_{\text{total}}}{\partial \alpha}\frac{\partial J}{\partial T} - \frac{\partial J}{\partial \alpha}\frac{\partial Q_{\text{total}}}{\partial T}}{
\frac{\partial J}{\partial \alpha} \frac{\partial Q_{\text{total}}}{\partial e^{w_+}} - \frac{\partial Q_{\text{total}}}{\partial \alpha}
\frac{\partial J}{\partial e^{w_+}}},
 \\ 
 \frac{\text{d}\alpha}{\text{d}T} =&~ \frac{
\frac{\partial Q_{\text{total}}}{\partial e^{w_+}}\frac{\partial J}{\partial T} - \frac{\partial J}{\partial e^{w_+}}\frac{\partial Q_{\text{total}}}{\partial T}}{
\frac{\partial J}{\partial e^{w_+}} \frac{\partial Q_{\text{total}}}{\partial \alpha} - \frac{\partial Q_{\text{total}}}{\partial e^{w_+}}
\frac{\partial J}{\partial \alpha}}\,.
    \end{aligned}
\end{equation}
Substituting by \eqref{thermodynamical quantities}, we arrive at
\begin{equation}\label{canonical ensemble}
    \begin{aligned}
        \frac{\text{d}e^{w_+}}{\text{d}T} &= \frac{4\pi \ell e^{w_+} \left(\alpha ^2 \cos ^2\theta_0 \left(2 \alpha ^2-20\pi \ell T+4\right)+\left(4 \alpha ^2-1\right) 4\pi\ell T+2\right)}{2 (4\pi \ell T-2) (4\pi\ell T-1) \left(\alpha ^2 \left(\alpha ^2+2\right) \cos^2\theta_0 +1\right)},\\
        \frac{\text{d}\alpha}{\text{d}T} &= -\frac{\alpha  \left(\alpha ^2+1\right) (4\pi\ell)^2 T \left(\alpha ^2 \cos ^2\theta_0+1\right)}{2 (4\pi\ell T-2) (4\pi\ell T-1) \left(\alpha ^2 \left(\alpha ^2+2\right) \cos^2\theta_0 +1\right)}.
    \end{aligned}
\end{equation}
Imposing these constraints is equivalent to asserting that $J$ and $Q_{\text{total}}$ are kept fixed. The heat capacity is then calculated using the formula
\begin{equation}
    C = T \left(\frac{\text{d}S}{\text{d}T}\right)_{J,Q_{\text{total}}},
\end{equation}
which can be expanded as
\begin{equation}
    C = T\left(\frac{\partial S}{\partial T} + \frac{\partial S}{\partial \alpha}\frac{\text{d} \alpha}{\text{d}T} + \frac{\partial S}{\partial e^{w_+}}\frac{\text{d} e^{w_+}}{\text{d}T}\right)_{J,Q_{\text{total}}}.
\end{equation}
Substituting by \eqref{entropy} and \eqref{canonical ensemble}, we end up with
\begin{equation}\begin{aligned}
C
=~&
    \frac{
\pi^{2}(1+\alpha^{2})\,\ell^{2}\,e^{w_{+}}\,T\,|\sin\theta_{0}|
}{
G_{3}\,(1-4\pi\ell T)\,(2-4\pi\ell T)\,
\sqrt{1+\alpha^{2}\cos^{2}\theta_{0}}
}
\\
&
\times
\left(
2
+
\frac{
4\pi\ell T\,
\big(
2\alpha^{2}
-1
-\alpha^{2}(\alpha^{2}+4)\cos^{2}\theta_{0}
\big)
}{
\alpha^{2}(\alpha^{2}+2)\cos^{2}\theta_{0}+1
}
\right)
\,.
\end{aligned}\end{equation}
Noting from the temperature formula \eqref{eq:hawkingtemperature} that $4\pi \ell T < 1$, it can be checked that the heat capacity is always positive, i.e., the black hole is locally stable.

The next natural step would be to go to the grand canonical ensemble where the rotation and charge are taken into account. However, in our case, this turns out to be trivial. To see this, notice that the grand canonical ensemble is defined by fixing $\Omega_H, T$ and $\Phi_{\text{el}}$. However, $\Omega_H$ only depends on $\alpha$ and upon inverting $\Omega_H(\alpha)$ (or,  equivalently, deriving $\alpha (\Omega_H)$), such that fixing $\Omega_H$ is equivalent to fixing $\alpha$. Having done so, we see that $\Phi_{\text{el}}$ is of the form $\Phi_{\text{el}}=f(\Omega_H)\sqrt{\frac{e^{w_-}}{e^{w_+}}}$, and from the expression for the temperature \eqref{eq:hawkingtemperature}, we can write $\Phi_{\text{el}} = f(\Omega_H)\sqrt{1-4\pi\ell T}$. Thus, $\Phi_{\text{el}}$ and $T$ are not independent variables when fixing $\Omega_H$ implying that the grand canonical ensemble, which assumes the independence of its fixed variables, is trivial.

\subsection{Symmetries of the geometry far and near}\label{sec:properties3d}
We derive the near-horizon geometry of the four-dimensional charged black hole and also consider the three-dimensional version. 
As pointed out, in both cases there exists an extremal limit when $e^{w_{-}}=e^{w_{+}}=: e^{w_{0}}$, as the temperature then vanishes \eqref{eq:hawkingtemperature}. We will approach extremality through
\begin{equation}
    e^{w_{\pm}}
    =
    e^{w_{0}}
    \pm
    2\pi \ell e^{w_{0}} T
    +
    \mathcal{O}(T^{2})
    \,,
\end{equation}
where $T$ is the Hawking temperature. 
To simultaneously zoom in onto the horizon we introduce a new radial coordinate $x$, which coincides with the horizon for $x=0$, an angular coordinate, and a rescaled time coordinate:
\begin{equation}
    e^{w}
    =
    e^{w_{+}}
    +
    4\pi \ell e^{w_{0}} T \sinh^{2}\frac{x}{2}
    \,,
    \quad
    \phi
    =
    \hat{\phi}
    -
    \frac{\alpha}{1+\alpha^{2}}
    \frac{\hat{t}}{2\pi \ell T}
    \,,
    \quad
    t
    =
    \frac{\hat{t}}{2\pi T}
    \,.
\end{equation}
Plugging this into the four-dimensional metric we find at leading order
\begin{equation}
    \text{d}s^{2}
    =
    \ell^{2}\frac{1+\alpha^{2}\cos^{2}\theta_{0}}{1+\alpha^{2}}
    \left(
        -\sinh^{2}x \text{d}\hat{t}^{2}
        +
        \text{d}x^{2}
        +
        (1+\alpha^{2})
        \text{d}\theta^{2}
        +
        \frac{(1+\alpha^{2})^{3}\sin^{2}\theta_{0}}{(1+\alpha^{2}\cos^{2}\theta_{0})^{2}}
        \text{d}\hat{\phi}^{2}
    \right)
    \,,
\end{equation}
which has the geometry AdS$_{2}\times \mathbb{R}\times S^{1}$.
In the case of the three-dimensional black hole we find that the near-horizon geometry becomes AdS$_{2}\times S^{1}$. For both geometries the Ricci scalar reads
\begin{equation}
    R=-\frac{2}{\ell^{2}}
    \frac{1+\alpha^{2}}{1+\alpha^{2}\cos^{2}\theta_{0}}
    \,,
\end{equation}
where $\alpha$ and $\theta_{0}$ can be interpreted as warping of the two-dimensional AdS length scale.

Finally, we consider the asymptotic symmetry group for the three-dimensional black hole. 
The first step is to recast the metric \eqref{eq:chargedmetric3d} into the Bondi gauge: consider first the rescaling $\bar{\phi} =  \ell \alpha \sin^2\theta_0 \phi$, then apply the coordinate transformations
\begin{equation}
    \text{d}t=\text{d}[u+X(w)]
    \,,
    \quad
    \text{d}\bar{\phi}=\text{d}[\tilde{\phi}+Y(w)]
    \,.
\end{equation}
Requiring $g_{w\tilde{\phi}}=g_{ww}=0$, yields
\begin{equation}
    \begin{aligned}
        X'(w) &= - \left(\frac{1+\alpha^2\cos^2\theta_0}{\alpha^2} \frac{1}{ f(w)} \right)Y'(w)\,,\\ 
        Y'(w) &= \frac{\ell\sqrt{1+\alpha^2\cos^2\theta_0}}{\sqrt{(1+\alpha^2) (f(w)-1)}}\frac{1}{\sqrt{-(\frac{1+\alpha^2\cos^2\theta_0}{\alpha^2 f(w)}+1)^2 + \frac{(1+\alpha^2)(1+\alpha^2\cos^2\theta_0)}{\alpha^4 f(w)} + \frac{1+\alpha^2}{\alpha^2}}}\,,
    \end{aligned}
\end{equation}
where $f(w)= m e^{-w}-Q^2e^{-2w}$. The metric in Bondi gauge is then written as
\begin{equation}
\begin{aligned}
    \text{d}s^2 =&~ 
    \left( -1+\frac{1+\alpha^2}{1+\alpha^2\cos^2\theta_0} f(w)\right) \text{d}u^2 + \frac{1+\alpha^2}{1+\alpha^2\cos^2\theta_0} f(w) \text{d}u\text{d}\tilde{\phi} 
    \\&+ \left(-2 X'(w) + 2 \frac{1+\alpha^2}{1+\alpha^2\cos^2\theta_0} f(w) (X'(w)+Y'(w))\right) \text{d}u\text{d}w 
    \\&
     + \left(\frac{1+\alpha^2}{\alpha^2} + \frac{1+\alpha^2}{1+\alpha^2\cos^2\theta_0} f(w)\right)\text{d}\tilde{\phi}^2\,.
    \end{aligned}
\end{equation}
First note that $\alpha\to 0$ is a singular limit, which is not surprising, because the geometry changes significantly in this limit, as it then partitions. In the limit $w \to \infty$, the metric can be expanded as
\begin{equation}
    \begin{aligned}
        \text{d}s^2 
        =&~ 
        (-1 + \mathcal{O}(e^{-w}))\text{d}u^2 
        + 
        \left(- 2 \ell \sqrt{\frac{1+\alpha^2\cos^2\theta_0}{1+\alpha^2}}
        + 
        \mathcal{O}(e^{-w})\right)\text{d}u\text{d}w 
        + 
        \mathcal{O}(e^{-w})\text{d}u\text{d}\tilde{\phi} 
        \\& 
        + 
        \left(\frac{1+\alpha^2}{\alpha^2} + \mathcal{O}(e^{-w})\right)\text{d}\tilde{\phi}^2. 
    \end{aligned}
\end{equation}
To calculate the asymptotic symmetry group, we write the metric as an expansion in $e^{-w}$ leaving the leading order to be varied depending on the boundary conditions. The metric form is
\begin{equation}
\begin{aligned}
    \text{d}s^2 
    =&~ 
    -(h_{uu} + \mathcal{O}(e^{-w}))\text{d}u^2 
    + 
    - 2 \ell \sqrt{\frac{1+\alpha^2\cos^2\theta_0}{1+\alpha^2}} (1+\mathcal{O}(e^{-w}))\text{d}u\text{d}w 
    \\& + 
    e^{-w}(h_{u\tilde{\phi}} + \mathcal{O}(e^{-w}))\text{d}u \text{d}\tilde{\phi} 
    + 
    \frac{1+\alpha^2}{\alpha^2}(1+\mathcal{O}(e^{-w}))\text{d}\tilde{\phi}^2.
    \end{aligned}
\end{equation}
    Thus, when calculating the generators of the asymptotic symmetry algebra, we have to take $\delta h_{u\tilde{\phi}} = 0$ since it is sub-leading and will correspond to local diffeomorphisms. Ignoring the $u\tilde{\phi}$ component of the defining equation of the generators: $\pounds_{\xi}g_{\mu\nu} = \delta g_{\mu\nu}$, we end up with the asymptotic symmetry group BMS$_2 \times U(1)$ which is more constraining than BMS$_3$ since super-translations are only radial and super-rotations can only be in the time direction. The asymptotic symmetry groups for asymptotically flat spacetimes with internal manifolds were studied in \cite{Ferko:2021bym}, however, in our case the internal circle is of fixed radius and the sub-leading $\text{d}u\text{d}\tilde{\phi}$ term is of the form $Ce^{-w}$ where $C$ is a constant. Thus, we have a more stringent asymptotic metric and consequently a more restricted asymptotic symmetry group featuring a rigid $U(1)$ rather than angle dependent transformations.
%
\subsection{The Penrose process}\label{sec:penrose}
%
The Penrose process was introduced in \cite{Penrose:1971uk} as a method to extract energy from rotating black holes. It can be phrased in the following way. Consider a particle with mass $M_p$ in the black hole geometry, its four-momentum is given by $p^{\mu} = M_p \frac{\text{d}x^{\mu}}{\text{d}\tau}$ where $\tau$ is the proper time. Its energy and angular momentum are conserved and are given by
\begin{equation}
    \begin{aligned}
        E &= - (\partial_t)_{\mu}p^{\mu} = -M_pg_{tt}\frac{\text{d}t}{\text{d}\tau} - M_p g_{t\phi}\frac{\text{d}\phi}{\text{d}\tau} ,\\
        P &= - (\partial_{\phi})_{\mu}p^{\mu} = -M_pg_{t\phi}\frac{\text{d}t}{\text{d}\tau} - M_p g_{\phi\phi}\frac{\text{d}\phi}{\text{d}\tau}.
    \end{aligned}
\end{equation}
Inside the ergoregion, the particle splits into two particles. One of them falls into the black hole and one escapes. Since in the ergoregion the timelike Killing vector is negative, it allows for particles with negative energy. The Penrose process is based on ensuring that the falling particle has negative energy, then by energy conservation, the escaping particle will have more energy than the initial particle. This energy is extracted from the rotational energy of the black hole, allowing us to harvest energy from the black hole in the form of kinetic energy.
For a Kerr black hole the theoretical maximum of energy to be extracted is a $0.29$ fraction of its mass \cite{Wald:1974kya,1972ApJ...178..347B}. We will show that for the three-dimensional geometry considered in this work, \eqref{eq:chargedmetric3d}, the extracted energy fraction ranges from $0$ to $1$, depending on the tuning parameters.
This can be understood as a remnant of the higher dimensional Penrose process whose efficiency can exceed $1$ \cite{Shaymatov:2024fle,Nozawa:2005eu}, which we will show from the large-$d$ perspective in Section \ref{sec:4}.\footnote{This is because there is no maximal angular momentum which is the energy to be extracted. We can, however, restrict ourselves to the ranges of constants that give less than a fraction of order $1$ efficiency. This shows us that this geometry is, in a way, more higher-dimensional than three-dimensional.}
To prevent the efficiency exceeding $1$, it turns out we have to require $\frac{\ell}{2 G_3}\leq 1$.

Let us now explicitly study the Penrose process. For the purpose of this subsection, we take $e^{w_-}=0$ (thus $Q=0$) since we are interested in extracting the rotational energy from the black hole. 
The particle's energy reads
\begin{equation}
	E 
	= 
	M_p 
	\left(
		1
		- 
		m \frac{1+\alpha^2}{1+\alpha^2\cos^2\theta_0}e^{-w}
	\right)
	\frac{\text{d}t}{\text{d}\tau} 
	- 
	\ell M_p \alpha  m \frac{1+\alpha^2}{1+\alpha^2\cos^2\theta_0}e^{-w}\frac{\text{d}\phi}{\text{d}\tau}\sin^2\theta_0
    \,,
\end{equation}
where $\tau$ is the proper time.
Since $E$ is negative inside the ergoregion, we can extract energy from the black hole via the Penrose process.

To calculate the maximum possible energy that can be extracted, we compute the irreducible mass \cite{Christodoulou:1970wf}. In three dimensions, the irreducible mass depends linearly on the area of the horizon, i.e., the irreducible mass depends linearly on the entropy: $M_{irr} = \frac{A_H}{16\pi G_3^2}$, where $A_H$ is the area of the horizon, but since $S = \frac{A_H}{4 G_3}$, we have $M_{irr} = \frac{S}{4\pi G_3}$. 
This represents the mass of the black hole when we extract all the possible energy that can be extracted. Using the results from section \ref{sec:charges}, we have the relation between mass and entropy at $e^{w_-}=0$:
\begin{equation}
    S = 2\pi \ell \sqrt{\frac{1+\alpha^2\cos^2\theta_0}{1+\alpha^2}}M|_{e^{w_-}=0}\,.
\end{equation}
The irreducible mass is then
\begin{equation}
    M_{irr} = \left(\frac{\ell}{ 2 G_3}\sqrt{\frac{1+\alpha^2\cos^2\theta_0}{1+\alpha^2}} \right)M|_{e^{w_-}=0}\,,
\end{equation}
thus, the extractable energy is given by
\begin{equation}\begin{aligned}
	\delta M 
	= &~
	M_{BH} - M_{irr}
	\\= &~ \left[
		1
		-
		\frac{\ell}{2 G_3}\sqrt{\frac{1+\alpha^2\cos^2\theta_0}{1+\alpha^2}}\right]
        M|_{e^{w_-}=0}
        \,.
\end{aligned}\end{equation}
The fraction of the extractable energy to the total energy is given by $
		1
		-
		\frac{\ell}{2 G_3}\sqrt{\frac{1+\alpha^2\cos^2\theta_0}{1+\alpha^2}}$.
 Depending on the values of $\alpha$, $\theta_0$, and the ratio $\frac{\ell}{G_3}$, the extractable fraction of energy can range from $0$ to $1$ of the total energy where $0$ and $1$ themselves are excluded. To see this, first choose $\theta_0 \approx 0$, so we can write $\cos^2\theta_0 = 1-\epsilon$ where $\epsilon \ll 1$. After using a Taylor expansion to only keep the linear order in $\epsilon$, the fraction of extractable energy (substituting in the extractable energy formula) becomes $1-\frac{\ell}{2 G_3}+\frac{\ell}{2 G_3}\frac{\epsilon \alpha^2}{2(1+\alpha^2)}$, which can be brought arbitrarily close to $1-\frac{\ell}{2 G_3}$ for any fixed $\alpha$. When $\frac{\ell}{2 G_3} \to 1$, the extractable energy fraction goes to $0$. A similar result holds for $\theta\approx\pi$. On the other hand, choosing $\theta_0 \approx \frac{\pi}{2}$ and writing $\cos^2\theta_0 = \epsilon$, the extracted energy in the limit $\epsilon \to 0$ is $ \frac{\ell}{2G_3} (1-\frac{1}{\sqrt{1+\alpha^2}})$ which can be arbitrarily close to $\frac{\ell}{2 G_3}$ if we have $\alpha \gg 1$. This in turn can be brought arbitrarily close to $1$ if $2G_3 \to \ell$, i.e., the more rotational energy the black hole has, the more efficiently we can extract it. 
 
 In summary, the Penrose process for this black hole is much more versatile than for conventional black holes, since there are many parameters which we can tune. For a fixed length scale $\frac{\ell}{2 G_3}\leq 1$, the maximum extractable energy fraction ranges from $(1-\frac{\ell}{2 G_3})$ for $\theta_0 \approx 0$ to $\frac{\ell}{2 G_3}$ at $\theta_0 \approx \frac{\pi}{2}$. The maximum range is when $\frac{\ell}{2G_3} \approx 1$. The range of maximum extractable energy ranges from an efficiency factor of $0$ to $1$. The most restrictive case is when $\ell=G_3$, when the maximum extractable energy ratio is $0.5$ regardless of the value of $\theta_0$.
\section{A large-$d$ perspective}\label{sec:4}
In this section we study how to take the large-$d$ limit of Myers-Perry black holes. In the case of a single rotational angle, we can derive the three- and four-dimensional solutions studied in the previous sections. This approach also offers a vantage point for interpreting quantities of the three- and four-dimensional models.
In Section \ref{sec:MP1} we revisit some important aspects of Myers-Perry black holes. We present careful considerations, such as zooming into the horizon, needed for taking the large-$d$ limit in Section \ref{sec:larged1}.
We reproduce the desired action in Section \ref{sec:reduction}.
\subsection{Myers-Perry black holes}\label{sec:MP1}
The Myers-Perry black hole \cite{Myers:1986un} is a solution of Einstein's equations in $d$ spacetime dimensions in vacuum. 
The solution can be considered as a higher-dimensional generalisation of the four-dimensional Kerr solution and reveals that beyond four dimensions, multiple independent rotational angles are possible, the maximum number of which is given by $n=\frac{d}{2}-1$ for even $d$, and $n=\frac{d}{2}-\frac{1}{2}$ for odd $d$. For even $d$, the metric for the $d$-dimensional Myers-Perry black hole is given by \cite{Myers:2011yc}
\begin{equation}\label{eq:even MP}
    \begin{aligned}
        \text{d}s^2 =&~ -\text{d}t^2 + \frac{\mu r}{\Pi F}\left(\text{d}t + \sum_{\substack{i=1}}^na_i\hat{\mu}_i^2\text{d}\hat{\phi}_i\right)^2 + \frac{\Pi F}{\Pi - \mu r}\text{d}r^2 \\&+ \sum_{\substack{i=1}}^n(r^2+a_i^2)\left(\text{d}\hat{\mu}_i^2 + \hat{\mu}_i^2\text{d}\hat{\phi}_i^2\right) + r^2\text{d}\hat{\psi}^2,
    \end{aligned}
\end{equation}
where
\begin{equation}
    \begin{aligned}
        F &= 1- \sum_{\substack{i=1}}^n\frac{a_i^2 \hat{\mu}_i^2}{r^2+a_i^2},\\
        \Pi &= \prod_{\substack{i=1}}^n(r^2+a_i^2).
    \end{aligned}
\end{equation}
For odd $d$, we have the metric
\begin{equation}\label{eq:odd MP}
    \begin{aligned}
        \text{d}s^2 =&~ -\text{d}t^2 + \frac{\mu r^2}{\Pi F}\left(\text{d}t + \sum_{\substack{i=1}}^na_i\hat{\mu}_i^2\text{d}\hat{\phi}_i\right)^2 + \frac{\Pi F}{\Pi - \mu r^2}\text{d}r^2 \\&+ \sum_{\substack{i=1}}^n(r^2+a_i^2)\left(\text{d}\hat{\mu}_i^2 + \hat{\mu}_i^2\text{d}\hat{\phi}_i^2\right),
    \end{aligned}
\end{equation}
where $F$ and $\Pi$ are the same as in the even $d$ metric. Notice that in the metrics the coordinates include $\hat{\mu}_i$ (and $\hat{\psi}$ for even $d$), which are the directional cosines of the transverse angular coordinates. Thus, they are related via $\sum_{i=1}^n \hat{\mu}_i^2=1$ for odd $d$ and $\sum_{i=1}^n \hat{\mu}_i^2 + \hat{\psi}^2=1$ for even $d$.

An important special case is formed by Myers-Perry black holes for $d>5$ with a single rotational angle since they are devoid of a maximal angular momentum, contrasting the existence of a Kerr bound on angular momentum in $d=4$ dimensions and Myers-Perry with two or more rotations. 
In this section, we are mainly interested in the large-$d$ limit of Myers-Perry black hole with a single rotational angle. The metric in this case is given by \cite{Myers:2011yc}
\begin{equation}\begin{aligned}\label{eq:line1}
    \text{d}s^2
    =& 
    -\text{d}t^2 + \frac{\mu}{r^{d-5}\Sigma}
    \left(
    	\text{d}t
	+
	a\sin^2\hat{\theta}\,\text{d}\hat{\phi}
	\right )^2
	+ 
	\frac{\Sigma}{\Delta}\text{d}r^2 
	+ 
	(r^2+a^2)\sin^2\hat{\theta}\,\text{d}\hat{\phi}^2 
    \\&
    	+ 
	\Sigma\, \text{d}\hat{\theta}^2 
    	+ 
	r^2\cos^2\hat{\theta}\,d\hat{\Omega}_{d-4}^2
    \,,
\end{aligned}\end{equation}
with
\begin{eqnarray}
	\Sigma 
	=& 
	r^2+a^2\cos^2\hat{\theta}\,,
	\\
	\Delta 
	=& 
	r^2 + a^2 - \frac{\mu}{r^{d-5}}\label{eq:warp1}
	\,,
\end{eqnarray}
where $\mu$ and $a$ are parameters that are proportional to the black hole mass and angular momentum, respectively.
In the introduced metric in \eqref{eq:line1} the angular coordinate $\hat{\theta}$ represents a polar coordinates and $\hat{\phi}$ runs over a circle, and $\hat{\Omega}_{d-4}$ represents a transverse $d-4$-sphere:
\begin{equation}
	\text{d}\hat{\Omega}^{2}_{d-4}
	=
	\sum^{d-4}_{i=1}
	\prod^{i-1}_{j=1}
	\sin^{2}\hat{\phi}_{j}\,
	\text{d}\hat{\phi}_{i}^{2}
	\,,
\end{equation}
where all $\phi_{j}$ angles run between $0$ and $2\pi$.
For $d=4$ we reproduce the Kerr metric in Boyer-Lindquist coordinates.

One way to infer the number of horizons is the following way.
The horizon(s) $r_h$ is determined by the real root(s) satisfying $\left.\Delta\right|_{r=r_h} = 0$. There is no non-trivial double (or higher degree) zero for $d\geq 5$. This is obvious for $d=5$, but in general we can see:
\begin{equation}
	\left.\Delta\right|_{r=r_\text{ext}} = 0
	\,,
	\quad
	\left.\partial_{r}\Delta\right|_{r=r_\text{ext}} = 0
	\,,
	\quad
	\Rightarrow
	\quad
	r_{\text{ext}}^{2}
	=
	\frac{5-d}{d-3}a^{2}
	=
	\left(
		\frac{2}{d-3}\frac{a^{2}}{\mu}
	\right)^{\frac{2}{5-d}}
	\,,
\end{equation}
which does not have real solutions for $d>5$ and as such does not allow for multiple horizons. 

We focus on Myers-Perry black holes with a single rotational angle. Their entropy, as usual, is given by $S = A/(4G_{d})$, where $A$ is the area of the horizon and $G_{d}$ the $d$ dimensional Newton's constant.
To define the temperature, we have to, firstly, define the angular velocity of the horizon evaluated on the outer horizon: $\Omega_H = - \frac{g_{t\phi}}{g_{\phi\phi}} = \frac{a}{r_H^2 + a^2}$, where $a$ is the rotation parameter and $r_H$ is the horizon's radius. Now, we can write down the vector generating the horizon
\begin{equation}
    \eta = \partial_t + \Omega_H \partial_{\phi}
    \,.
\end{equation}
The surface gravity $\kappa$ of the horizon is given by the equation 
\begin{equation}
    \eta^{\mu}\nabla_{\mu}\eta^{\nu} = \kappa \eta^{\nu}.
\end{equation}
The resulting Hawking temperature is given by
\begin{equation}\label{eq:surfgrav}
	T = \frac{\kappa}{2\pi}
	\,.
\end{equation}
These thermodynamical quantities satisfy the Smarr formula
\begin{equation}\label{eq:smarr1}
    \frac{d-3}{d-2}M = \Omega_H J + T S
    \,,
\end{equation}
where $M$ and $J$ are the mass and angular momentum of the black hole calculated via the Komar integral to be $M = \frac{(d-2)\Omega_{d-2}}{16 \pi G_{d}}\mu$ where $\Omega_{d-2}$ is the area of $S^{d-2}$, and 
\begin{equation}\label{eq:rotandmass}
    J = \frac{2a}{d-2}M
    \,.
\end{equation}
See \cite{Myers:2011yc} for more details on the Myers-Perry solution.

%
\subsection{Large-$d$ geometry of Myers-Perry}\label{sec:larged1}
%
In the previous subsection we reviewed relevant aspects of the Myers-Perry solution. 
We will now prepare coordinates and conventions in a manner that allows for taking the large-$d$ limit whilst retaining a non-trivial solution.
Viewing the large-$d$ limit as a decoupling limit, in the same spirit as, e.g., taking the extremal limit for charged black hole, our choice will allow for a non-trivial near-horizon description of the decoupling limit.

In order to obtain a non-trivial near-horizon description in the large-$d$ limit, we introduce the parametrisation $\mu=r_{0}^{d-3}$, where $r_{0}$ is a length scale. From $\Delta(r_{h})=0$, see \eqref{eq:warp1}, we establish
\begin{equation}\begin{aligned}
	\left(\frac{r_{0}}{r_{h}}\right)^{d-3}
	=&~
	1+\alpha^{2}
	\,,
	\\
	\frac{r_{0}}{r_{h}}
	=&~
	1
	+
	\mathcal{O}(d^{-1})
	\,,
\end{aligned}\end{equation}
where we introduced the dimensionless quantity
\begin{equation}\label{eq:defalpha}
	\alpha=\frac{a}{r_{h}}\geq0
	\,.
\end{equation}
In order to analyse the situation when we take the large-$d$ limit, we introduce the dimensionless coordinate $w$:
\begin{equation}
	\frac{r}{r_{h}}
	=
	1
	+
	\frac{w}{d}
	\,,
\end{equation}
where $w=0$ yields the location of the outer horizon. To capture the black hole dynamics, we take $w\lesssim d^{n}$ where for consistency with taking the large-$d$ limit we consider the range $0\leq n<1$, see \cite{Sybesma:2022nby} for a discussion. After taking the large-$d$ limit $-\infty<w<\infty$, we find that
\begin{equation}\begin{aligned}
	\frac{r}{r_{h}}
	\stackrel{d\to \infty}{=}&~
	1
	\,,
	\\
	\left(\frac{r}{r_{h}}\right)^{d}
	\stackrel{d\to \infty}{=}&~
	e^{w}
	\,,
	\\
	\text{d}r
	\stackrel{d\to \infty}{=}&~
	\frac{r_{h}}{d}
	~\text{d}w
	\,,
\end{aligned}\end{equation}
where we keep the following quotient fixed
\begin{equation}
	\ell\stackrel{d\to \infty}{=}\frac{r_{h}}{d}
	\,.
\end{equation}
We can in particular consider what happens to different non-trivial terms in the line element of \eqref{eq:line1} when we take the large-$d$ limit:
\begin{equation}\begin{aligned}
	\frac{\Delta}{r_{h}^{2}}
	\stackrel{d\to \infty}{=}&~
	(1+\alpha^{2})
	(1-e^{-w})	
	\,,
	\\
	\frac{\Sigma}{r_{h}^{2}}
	\stackrel{d\to \infty}{=}&~
	1
	+
	\alpha^{2}\cos^{2}\theta_0
	\,,
	\\
	\frac{\mu}{r^{d-5}\Sigma}
	\stackrel{d\to \infty}{=}&~
	e^{-w}\frac{1+\alpha^{2}}{1+\alpha^{2}\cos^{2}\theta_0}
	\,,
\end{aligned}\end{equation}
where $\theta_{0}$ is defined in the following way.
Simultaneous with the large-$d$ limit, we also aim to zoom in into around the angular values denoted with subscript $0$:
\begin{equation}\begin{aligned}\label{eq:deftheta0}
	\hat{\theta}
	=&~
	\theta_{0}
	+
	\frac{\theta}{d}
	\,,
	\\
	\hat{\phi}
	=&~
	\phi_{0}
	+
	\frac{\phi}{d}
	\,,
	\\
	\hat{\Omega}_{d-4}
	=&~
	\Omega_{0,d-4}
	+
	\frac{\Omega_{d-4}}{d}
	\,.
\end{aligned}\end{equation}
The condensed notation $\Omega_{0,d-4}$ and $\Omega_{d-4}$ contains $\phi_{0,j}$ and $\phi_{j}$ in an appropriate manner. We note that in principle $0\leq \theta_{0}\leq\pi$ and the other $0$ subscripted quantities in a similar fashion.
The unhatted symbols carrying no zero subscript, parametrise the zoomed in region and run over the real numbers when the large-$d$ limit is taken. In particular, when taking the large-$d$ limit:
\begin{equation}\begin{aligned}
	\text{if}
	\quad
	0<\theta_{0}<\pi
	\quad
	\text{then}&
	\quad
	-\infty<\theta<\infty
	\,,
	\\
	\text{if}
	\quad
	\theta_{0}=0
	\quad
	\text{then}&
	\quad
	0<\theta<\infty
	\,,
	\\
	\text{if}
	\quad
	\theta_{0}=\pi
	\quad
	\text{then}&
	\quad
	-\infty<\theta<0
	\,.
\end{aligned}\end{equation}
In contrast, we do choose $\phi\in[0,2\pi]$, so that we can keep angular momentum when taking the large-$d$ limit. 
From the perspective of the three-dimensional reduced model, we alternatively could have opted for integrating out $\theta$, without any loss of generality.

Using the established conventions and taking the large-$d$ limit of the Myers-Perry metric \eqref{eq:line1}, we get
\cite{Andrade:2018nsz}:
\begin{equation}\label{eq:largedmetric}
\begin{aligned}
    \text{d}s^2 
    =& 
    -\text{d}t^2 
    + 
    \frac{1+\alpha^2}{1+\alpha^2\cos^2\theta_0}
    e^{-w}(\text{d}t + \ell \alpha \sin^2\theta_0\,\text{d}\phi)^2 
    + 
    \ell^2
    \frac{ \text{d}w^2}{F(w)}
     + 
     \ell^2 (1+\alpha^2)\sin^2\theta_0\,\text{d}\phi^2
    \\ & 
    + 
    \ell^2(1+\alpha^2\cos^2\theta_0)\text{d}\theta^2 
    + 
    \ell^2\cos^2\theta_0\,\text{d}\Omega_{d-4}^2
    \,,
    \end{aligned}
\end{equation}
where $F(w)$ is an emblackening factor. Before analysing the emblackening factor, we first inspect the structure of the metric. In principle, the part of the metric involving $\text{d}t$, $\text{d}\phi$ and $\text{d}w$ spans a non-trivial geometric. This three-dimensional part of the geometry is supplied with a $\text{d}\theta$ line and a large-$d$ spherical part, which each factorise from each other. Henceforth, we will be interested in studying consistent truncations of the metric above, after compactifying the large-$d$ spherical part. 
On top of that, we will also consider the cases where we do (the three-dimensional case) and do not (the four-dimensional case) integrate out the $\theta$ direction. Meanwhile, in particular the values of $\alpha$ (defined in \eqref{eq:defalpha}) and $\theta_{0}$ (defined in \eqref{eq:deftheta0}) have an important impact on the geometry, which we will touch upon after analysing the emblackening factor $F(w)$:
\begin{equation}
	F(w) = \frac{1+\alpha^2}{1+\alpha^2\cos^2\theta_0}(1-e^{-w})
	\,.
\end{equation} 
We note that for $w=0$ the emblackening factor $F$ switches sign. In particular for $w\to\infty$ we find that $F\to\text{constant}$ and that $F\to-\infty$ when $w\to-\infty$. 
To connect to the behaviour of the emblackening factor we study the Ricci scalar $R$. 
The Ricci scalar for this geometry is
\begin{equation}
    R = \frac{1+\alpha^2}{2+\alpha^2 + \alpha^2 \cos^2\theta_0}\frac{e^{-w}}{\ell^2}
    \,.
\end{equation}
For $w\to\infty$ we recover flat space, as $R\to 0$. There is a curvature singularity at $w\to-\infty$ where the Ricci scalar blows up and $R$ is finite at the horizon $w=0$. 
Piecing together the evidence above we claim that the metric in \eqref{eq:largedmetric} can describe an asymptotically flat space black hole with the horizon at $w=0$.
The black hole has particular features depending on the values of $\alpha$ and $\theta_{0}$. In particular, for $\theta_{0}=0,\pi$ or $\alpha=0$ we lose rotation but find a black hole solution that is closely related to the two dimensional black hole of string theory. Indeed, in that case we can just view the solution as the aforementioned two-dimensional Witten black hole lifted to higher dimensions. When $0<\theta_{0}<\pi$ and $\alpha\neq0$, we find a rotating version of the Witten black hole, that needs a third direction, the $\phi$ direction, to support its rotation. 

Finally, setting $Q=0$ and $m=1$ in \eqref{eq:chargedmetric} recovers the relevant part of the higher $d$ metric in \eqref{eq:largedmetric}, making it a consistent truncation coming from the large-$d$ theory.

\subsection{Reduction at the level of the action}\label{sec:reduction}
In this section we will discuss a four-dimensional reduction of the large-$d$ geometry presented in \eqref{eq:largedmetric}. This amounts to reducing the transverse $d-4$-sphere. We also construct the corresponding action. 
With the Myers-Perry solution \eqref{eq:line1} in the back of our mind we write the following reduction Ansatz:
\begin{equation}
	\text{d}s^{2}
	=
	\gamma_{\mu\nu}\text{d}x^{\mu}\text{d}x^{\nu}
	+
	r_{h}^{2}e^{-\phi(r)\frac{4}{d}}
	\tilde{\psi}^{2}(\theta)
	\text{d}\Omega^{2}_{d-4}
	\,,
\end{equation}
\begin{equation}
	\tilde{\psi}(\theta)
	=
	\cos\left(
		\theta_{0}+\frac{2}{\tan\theta_{0}}\frac{\psi(\theta)}{d}
	\right)
	\,,
\end{equation}
where $\mu$ and $\nu$ run over $t$, $\theta$, $r(w)$, $\phi$. We note that we introduced $\tan\theta_{0}$ as a convenient rescaling, which means that we will exclude the values $\theta_{0}=\pi/2$ in addition to $\theta_{0}=0,\pi$, for which one would have to perform a separate analysis.
We note that in principle reading off from the solution we expect the solutions
\begin{equation}\label{eq:consistentsolution}
	\phi(r)
	=
	-\frac{1}{2}w
	\,,
	\quad
	\psi(\theta)
	=
	\frac{\tan\theta_{0}}{2}\theta
	\,,
\end{equation}
but which we will not invoke at this point.

Let us consider what the volume becomes
\begin{equation}\begin{aligned}
	\Pi^{d-2}_{i=1}\int^{2\pi}_{0}\text{d}\phi_{i}
	\sqrt{-g}
	=~~&~
	\Pi^{d-2}_{i=1}\int^{2\pi}_{0}\text{d}\phi_{i}
	\sqrt{-\gamma}
	r_{h}^{d-2}
	(\cos \theta_{0})^{d-2}
	e^{-2\frac{d-2}{d}\phi }
	\left[
		\frac{
		\tilde{\psi}	
		}{
			\cos \theta_{0}
		}
	\right]^{d-2}
	\\
	\stackrel{d\to \infty}{=}&~
	\left[
	\Pi^{d-2}_{i=1}\int^{2\pi}_{0}\text{d}\phi_{i}
	\sqrt{-\gamma}
	r_{h}^{d-2}
	(\cos \theta_{0})^{d-2}
	\right]
	e^{-2 \phi }
	e^{-2 \psi }
	\,,
\end{aligned}\end{equation}
where we assumed that $\gamma_{\mu\nu}$ does not depend on the angles of the transverse $d-4$-sphere. The part in between square brackets in the last line will be absorbed into Newton's constant as an overall factor in the action. For the Ricci scalar we find
\begin{equation}\begin{aligned}
	R
	=&~
	R_{\gamma}
	+
	4\nabla_{\gamma}^{2}\phi
	-
	\frac{d+1}{d} 4(\nabla_{\gamma}\phi)^{2}
	-
	2d\frac{\nabla_{\gamma}^{2}\tilde{\psi}}{\tilde{\psi}}
	-
	d(d-1)\frac{(\nabla_{\gamma}\tilde{\psi})^{2}}{\tilde{\psi}^{2}}
	\\&
	+
	4(d+1) \frac{ (\nabla_{\gamma})_{\mu} \phi(\nabla_{\gamma})^{\mu}\tilde{\psi}}{\tilde{\psi}}
	+
	d(d-1)\frac{e^{\frac{4}{d} \phi}}{r_{h}^{2}\tilde{\psi}^{2}}
	\,,
\end{aligned}\end{equation}
where the Ricci scalar $R_{\gamma}$ and covariant derivatives $\nabla_{\gamma}$ are with respect to the dimensionally reduced metric $\gamma_{\mu\nu}$.
At large $d$ we arrive at
\begin{equation}
	R
	\stackrel{d\to \infty}{=}
	R_{\gamma}
	+
	4\nabla_{\gamma}^{2}\phi
	-
	4(\nabla_{\gamma}\phi)^{2}
	+
	4(\nabla_{\gamma})^{2}\psi
	-
	4(\nabla_{\gamma}\psi)^{2}
	+
	8(\nabla_{\gamma})_{\mu} \phi(\nabla_{\gamma})^{\mu}\psi
	+
	\frac{1}{\ell^{2}\cos^{2}\theta_{0}}
	\,.
\end{equation}
Plugging this into to the vacuum $d$-dimensional Einstein-Hilbert action, performing integration by parts, and taking the large-$d$ limit, we obtain
\begin{equation}
	S_{4d}
	=
	\int \text{d}^{4}x
    \frac{
        \sqrt{-\gamma}~e^{-2\phi-2\psi}
    }{16\pi G_{4}}
	\left[
		R_{\gamma}
		+
		4(\nabla_{\gamma}\phi)^{2}
		+
		4(\nabla_{\gamma}\psi)^{2}
		+
		8(\nabla_{\gamma})_{\mu} \phi(\nabla_{\gamma})^{\mu}\psi
		+
		\frac{1}{\ell^{2}\cos^{2}\theta_{0}}
	\right]
	\,,
\end{equation}
where all prefactors are absorbed into $G_{4}$: the four-dimensional Newton's constant, $\frac{1}{G_{4}}:=\lim_{d\to\infty}\frac{1}{G_{d}}\left[
	\Pi^{d-2}_{i=1}\int^{2\pi}_{0}\text{d}\phi_{i}
	\sqrt{-\gamma}
	r_{h}^{d-2}
	(\cos \theta_{0})^{d-2}
	\right]$\,. This resulting four-dimensional action can be rewritten using the following field redefinition
\begin{equation}
	\phi(w)
	+
	\psi(\theta)
	=
	\Phi(w,\theta)
	\,,
\end{equation} 
which yields
\begin{equation}\label{eq:4dcghs}
	S_{4d}
	=
	\frac{1}{16\pi G_{4}}
	\int \text{d}^{4}x\sqrt{-g}~e^{-2\Phi}
	\left[
		R
		+
		4(\nabla\Phi)^{2}
		+
		4\lambda^{2}
	\right]
	\,,
\end{equation}
where we introduced
\begin{equation}
	4\lambda^{2}
	=
	\frac{1}{\ell^{2}\cos^{2}\theta_{0}}
	\,,
\end{equation}
and we dropped $\gamma$ in favour of $g$ as notation for the metric.
We have now derived the four-dimensional action and corresponding solution we presented in Section \ref{sec:2}. There we showed how the three-dimensional setup follows from the four-dimensional one.

\section{A black hole possessing multi angular momentum-like charges}\label{newsec5}
Generalising the large-$d$ tools constructed in the previous paper we can derive a solution to the general $D$-dimensional low-energy effective action \eqref{eq:eff_intro} that possesses multi angular momentum-like charges. 
The Einstein's equation and dilaton equation of motion associated to \eqref{eq:eff_intro} are:
\begin{equation}
    \begin{aligned}
         R_{\mu\nu} 
    - 
    \frac{1}{2}g_{\mu\nu}R 
    + 
    2 g_{\mu\nu}(\partial \Phi)^2 
    + 
    2\nabla_{\mu}\nabla_{\nu}\Phi 
    - 
    2g_{\mu\nu} \Box \Phi 
    - 
    2\lambda^{2}g_{\mu\nu} 
    = 0
    \,,
    \end{aligned}
\end{equation}
\begin{equation}
    \begin{aligned}
        R 
    - 
    4(\partial \Phi)^2 
    + 
   4\lambda^{2}
    + 
    4 \Box \Phi =0
    \,.
    \end{aligned}
\end{equation}
We claim that for any dimension $D\geq 2$, the equations of motion written above are solved by the following geometry\footnote{A Mathematica notebook with more details is available upon request.}
\begin{equation}\label{sol.any d}
    \begin{aligned}
        \text{d}s^2_D =&~ -\text{d}t^2+ \mathcal{A}_D(1-\mathcal{F})\left(\text{d}t + \sum_{\mathclap{i=1}}^{D-2}\ell \alpha_i \mu_0{}_i^2 \text{d}\phi_i\right)^2 + \frac{\ell^2 \text{d}w^2}{\mathcal{A}_D \mathcal{F}} 
        \\&~ 
        +\ell^2 \sum_{\mathclap{i=1}}^{D-2}(1+\alpha_i^2)\mu_0{}_i^2 \text{d}\phi_i^2 
        \,,
    \end{aligned}
\end{equation}
with
\begin{equation}
    \mathcal{F}
    =
    1-m e^{-w}
    \,,
    \quad 
    4\lambda^2= \frac{\mathcal{A}_D}{\ell^2}
    \,,
    \quad
     \mathcal{A}_D = 
    \left(1-\sum^{D-2}_{i=1}\frac{\alpha_{i}^{2}\mu_0{}_i^2}{1+\alpha^{2}_{i}}
    \right)^{-1}
    \,,
\end{equation}
and supplemented with a dilaton solution that also exhibits an asymptotically linear dilaton vacuum
\begin{equation}
    \Phi
    =
    -\frac{1}{2}w
    \,.
\end{equation}
In the above solution $\alpha_{i}$ are the parameters proportional to the angular momenta associated to the angles $\phi_{i}$. 
The parameters $\mu_{0}{}_{i}>0$ encode the relative warping of the asymptotic spacetime.
The Ricci scalar is given by
\begin{equation}
    R = \frac{\mathcal{A}_D}{\ell^2}e^{-w}
    \,,
\end{equation}
which reveals a curvature singularity for $w\to-\infty$ and an asymptotic flat space behaviour.

Due to a similar emblackening factor as the single-angle rotating case the thermodynamic behaviour is qualitatively the same, notably exhibiting no Kerr-like bound.
Its asymptotic symmetries are
\begin{equation}
\text{BMS}_2 \times \underbrace{U(1) \times \cdots \times U(1)}_{D-2 \ \text{times}}
\,,
\end{equation}
which is more stringent than the full BMS$_D$ group.
Finally, contrasting the Myers-Perry solution that has $\lfloor(D-1)/2\rfloor$ angular momenta, the here treated solution allows for $D-2$ angular momenta-like charges.

\section{Discussion}\label{sec:5}
In this article we presented new rotating black hole solutions to the low-energy effective action of string theory, which can be viewed as a model of dilaton gravity. We focussed on a three-dimensional model as well as its four-dimensional extension. We find that its thermodynamics exhibit a mixture of higher dimensional Myers-Perry qualities such as having no extremal bound associated to rotation and two-dimensional Witten black hole features such as a temperature that does not depend on the black hole mass. The tool we used to track down the above solutions was studying the large-$d$ behaviour of Myers-Perry black holes.
We furthermore also showed the extensions of the solutions for when we add an electric charge or extra angular momentum-like charges for arbitrary dimensions.
Finally, we point out that all of the here presented models allow for a linear dilaton vacuum and the asymptotic symmetries of these models turn out to be more stringent than their Einstein gravity equivalents.

As an outlook to future work, we recall that the Witten black hole, which the here studied solutions closely mirror, has for example been important for studying dualities in string theory \cite{Fateev:2000ik,Alvarez:1993qi}, studying connections to little string theory \cite{Aharony:1998ub,Maldacena:1997cg}, evaporation \cite{Callan:1992rs,Russo:1992ax}, the Page curve \cite{Gautason:2020tmk}, and holographic complexity \cite{schneiderbauer2020semi,Schneiderbauer:2020isp}. It would be interesting to further explore our solutions along those lines. 
More along the direction of studying Myers-Perry black holes, it can be insightful to use our results to, e.g., study black holes with different topologies and instabilities such as the Gregory-Laflamme type \cite{Gregory:1993vy}.
In particular a further study of the here presented black holes possessing multi angular momentum-like charges could reveal insights into dynamical features such as (in-)stability.

Finally, we want to end with the observation that the way we used the large-$d$ limit in this article is very much in the same spirit as using a the near-extremal limit to generate solutions and models with a large degree of analytic control. We are curious to learn what else this limit can uncover in the future.
%
\acknowledgments
We thank Vyshnav Mohan, Bo Sundborg, and L\'{a}rus Thorlacius for stimulating discussions in the context of this article. We want to especially thank Roberto Emparan for suggesting the current narrative of this work and insightful comments. We also thank Gerben Oling for valuable comments on the draft. 
The work of WS is supported by a Starting Grant 2023-03373 from the Swedish Research Council and the Olle Engkvists Stiftelse.
PT is supported by a Nordita visiting PhD student fellowship, Primus grant PRIMUS/23/SCI/005 and GAUK 425425 from Charles University.

\appendix
%
\section{Alternative derivation for rotation and temperature}\label{appendixA}
We assume the following near-horizon form:
\begin{equation}\begin{aligned}
	\text{d}s^{2}
	=&
	-\epsilon^{2}x^{2}
	\left(
		a_{t}\text{d}t
		-
		a_{\phi}\text{d}\phi
	\right)^{2}
	+
	\epsilon^{2}\text{d}x^{2}
	+
	\epsilon^{2}
	\left(
		b_{t}\text{d}t
		-
		b_{\phi}\text{d}\phi
	\right)^{2}
	\\=&
    \epsilon^{2}
    \left[
	   -x^{2}\text{d}\tilde{t}^{2}
	   +
       \text{d}\tilde{x}^{2}
	   +
	   \text{d}\tilde{\phi}^{2}
    \right]
	\,,
\end{aligned}\end{equation}
where
\begin{equation}
	\tilde{t}
	=
	a_{t}t
	-
	a_{\phi}\phi
	\,,
	\quad
	\tilde{\phi}
	=
	b_{t}t
	-
	b_{\phi}\phi
	\,.
\end{equation}
For regularity we now impose
\begin{equation}
	(\tilde{t},\tilde{\phi})
    \sim
    (\tilde{t}+i2\pi,\tilde{\phi}+i2\pi)
	\,.
\end{equation}
Regularity allows us to read off the following quantities (from some thermal partition function argument):
\begin{equation}
	(t,\phi)\sim(t+i\beta,\phi-i\beta\Omega)
	\,,
\end{equation}
which yields
\begin{equation}
    2\pi = \beta(a_{t}+a_{\phi}\Omega) 
    \,,
    \quad
    2\pi = \beta(b_{t}+b_{\phi}\Omega)
    \,.
\end{equation}
Rearranging yields:
\begin{equation}
	\beta=2\pi\lim_{\epsilon\to0}\frac{a_{\phi}-b_{\phi} }{a_{\phi} b_{t}-a_{t} b_{\phi}}
	\,,
	\quad
	\Omega=\lim_{\epsilon\to0}\frac{b_{t} -a_{t}}{a_{\phi} - b_{\phi}}
	\,.
\end{equation}
We take the $\epsilon\to0$ limit because we are interested in the angular velocity at the outer horizon. 

In the case of metric \eqref{eq:chargedmetric3d} and definition \eqref{eq:wplus} we establish 
\begin{equation}\begin{aligned}
	a_{t}
    =
    -
    \frac{
        e^{w_{-}}-e^{w_{+}}
    }{
        e^{w_{+}}
    }
    \frac{
        1+\alpha^{2}
    }{
        2\ell(1+\alpha^{2}\cos^{2}\theta_{0})
    }
    \,,
    \quad
    a_{\phi}
    =
    -\alpha\sin^{2}\theta_{0}\ell
	\,,
\end{aligned}\end{equation}	
\begin{equation}
    b_{t}
    =
    \frac{\alpha \sin\theta_{0}}{
        \epsilon
        \sqrt{
            1
            +
            \alpha^{2}\cos^{2}\theta_{0}
        }
    }
    \,,
    \quad
    b_{\phi}
    =
    -\frac{1+\alpha^{2}}{\alpha}\ell b_{t}
    \,.
\end{equation}
Plugging into the formula above yields
\begin{equation}
	T
	=
    \frac{e^{w_{+}}-e^{w_{-}}}{e^{w_{+}}}
	\frac{1}{4\pi \ell}
	\,,
	\quad
	\Omega
	=
	\frac{\alpha}{1+\alpha^{2}}\frac{1}{\ell}
	\,.
\end{equation}
In the uncharged case, the temperature is independent of $\alpha$, which is related to the angular velocity of this black hole, and the black hole mass as well. This last fact is reminiscent of the two-dimensional Witten black hole \cite{Mandal:1991tz,Witten:1991yr,Callan:1992rs}.

\bibliographystyle{JHEP}
\bibliography{refds}

@article{Fateev:2000ik,
    author = "Fateev, V. and Zamolodchikov, Alexander B. and Zamolodchikov, Alexei B.",
    title = "{Boundary Liouville field theory. 1. Boundary state and boundary two point function}",
    eprint = "hep-th/0001012",
    archivePrefix = "arXiv",
    reportNumber = "RUNHETC-2000-01",
    month = "1",
    year = "2000"
}

@article{Harksen:2024uik,
    author = "Harksen, Matthias and Sybesma, Watse",
    title = "{The spectrum of a quantum Lifshitz black hole in two dimensions}",
    eprint = "2408.15336",
    archivePrefix = "arXiv",
    primaryClass = "hep-th",
    doi = "10.1007/JHEP02(2025)214",
    journal = "JHEP",
    volume = "02",
    pages = "214",
    year = "2025"
}

@article{Gregory:1993vy,
    author = "Gregory, R. and Laflamme, R.",
    title = "{Black strings and p-branes are unstable}",
    eprint = "hep-th/9301052",
    archivePrefix = "arXiv",
    doi = "10.1103/PhysRevLett.70.2837",
    journal = "Phys. Rev. Lett.",
    volume = "70",
    pages = "2837--2840",
    year = "1993"
}

@article{Alvarez:1993qi,
    author = "Alvarez, E. and Alvarez-Gaume, Luis and Barbon, J. L. F. and Lozano, Y.",
    title = "{Some global aspects of duality in string theory}",
    eprint = "hep-th/9309039",
    archivePrefix = "arXiv",
    reportNumber = "CERN-TH-6991-93, FTUAM-93-28",
    doi = "10.1016/0550-3213(94)90067-1",
    journal = "Nucl. Phys. B",
    volume = "415",
    pages = "71--100",
    year = "1994"
}

@article{Green:1984sg,
    author = "Green, Michael B. and Schwarz, John H.",
    title = "{Anomaly Cancellation in Supersymmetric D=10 Gauge Theory and Superstring Theory}",
    reportNumber = "CALT-68-1182",
    doi = "10.1016/0370-2693(84)91565-X",
    journal = "Phys. Lett. B",
    volume = "149",
    pages = "117--122",
    year = "1984"
}

@article{Callan:1985ia,
    author = "Callan, Jr., Curtis G. and Martinec, E. J. and Perry, M. J. and Friedan, D.",
    title = "{Strings in Background Fields}",
    reportNumber = "PRINT-85-0734 (PRINCETON)",
    doi = "10.1016/0550-3213(85)90506-1",
    journal = "Nucl. Phys. B",
    volume = "262",
    pages = "593--609",
    year = "1985"
}

@article{Fradkin:1984pq,
    author = "Fradkin, E. S. and Tseytlin, Arkady A.",
    title = "{Effective Field Theory from Quantized Strings}",
    reportNumber = "LEBEDEV-84-261",
    doi = "10.1016/0370-2693(85)91190-6",
    journal = "Phys. Lett. B",
    volume = "158",
    pages = "316--322",
    year = "1985"
}

@article{Giveon:1994fu,
    author = "Giveon, Amit and Porrati, Massimo and Rabinovici, Eliezer",
    title = "{Target space duality in string theory}",
    eprint = "hep-th/9401139",
    archivePrefix = "arXiv",
    reportNumber = "RI-1-94, NYU-TH-94-01-01",
    doi = "10.1016/0370-1573(94)90070-1",
    journal = "Phys. Rept.",
    volume = "244",
    pages = "77--202",
    year = "1994"
}

@article{Gibbons:1994vm,
    author = "Gibbons, G. W. and Horowitz, Gary T. and Townsend, P. K.",
    title = "{Higher dimensional resolution of dilatonic black hole singularities}",
    eprint = "hep-th/9410073",
    archivePrefix = "arXiv",
    reportNumber = "DAMTP-R-94-28, UCSBTH-94-35",
    doi = "10.1088/0264-9381/12/2/004",
    journal = "Class. Quant. Grav.",
    volume = "12",
    pages = "297--318",
    year = "1995"
}

@article{Fraiman:2018ebo,
    author = "Fraiman, Bernardo and Gra{\~n}a, Mariana and N{\'u}{\~n}ez, Carmen A.",
    title = "{A new twist on heterotic string compactifications}",
    eprint = "1805.11128",
    archivePrefix = "arXiv",
    primaryClass = "hep-th",
    doi = "10.1007/JHEP09(2018)078",
    journal = "JHEP",
    volume = "09",
    pages = "078",
    year = "2018"
}

@article{PhysRevD.35.648,
  title = {On toroidal compactification of heterotic superstrings},
  author = {Ginsparg, Paul},
  journal = {Phys. Rev. D},
  volume = {35},
  issue = {2},
  pages = {648--654},
  numpages = {0},
  year = {1987},
  month = {Jan},
  publisher = {American Physical Society},
  doi = {10.1103/PhysRevD.35.648},
  url = {https://link.aps.org/doi/10.1103/PhysRevD.35.648}
}

@article{Metsaev:1998it,
    author = "Metsaev, R. R. and Tseytlin, Arkady A.",
    title = "{Type IIB superstring action in AdS(5) x S**5 background}",
    eprint = "hep-th/9805028",
    archivePrefix = "arXiv",
    reportNumber = "FIAN-TD-98-21, IMPERIAL-TP-97-98-44, NSF-ITP-98-055",
    doi = "10.1016/S0550-3213(98)00570-7",
    journal = "Nucl. Phys. B",
    volume = "533",
    pages = "109--126",
    year = "1998"
}

@article{Hubsch:1986ny,
    author = "Hubsch, Tristan",
    title = "{Calabi-yau Manifolds: Motivations and Constructions}",
    reportNumber = "MDDP-PP-86-149",
    doi = "10.1007/BF01210616",
    journal = "Commun. Math. Phys.",
    volume = "108",
    pages = "291",
    year = "1987"
}

@article{Choi:2004vb,
    author = "Choi, Kang-Sin",
    title = "{Spectra of heterotic strings on orbifolds}",
    eprint = "hep-th/0405195",
    archivePrefix = "arXiv",
    reportNumber = "SNUTP-04-011",
    doi = "10.1016/j.nuclphysb.2004.11.045",
    journal = "Nucl. Phys. B",
    volume = "708",
    pages = "194--214",
    year = "2005"
}

@article{Dixon:1985jw,
    author = "Dixon, Lance J. and Harvey, Jeffrey A. and Vafa, C. and Witten, Edward",
    editor = "Schellekens, B.",
    title = "{Strings on Orbifolds}",
    reportNumber = "PRINT-85-0616 (PRINCETON)",
    doi = "10.1016/0550-3213(85)90593-0",
    journal = "Nucl. Phys. B",
    volume = "261",
    pages = "678--686",
    year = "1985"
}

@article{DIXON1986285,
title = {Strings on orbifolds (II)},
journal = {Nuclear Physics B},
volume = {274},
number = {2},
pages = {285-314},
year = {1986},
issn = {0550-3213},
doi = {https://doi.org/10.1016/0550-3213(86)90287-7},
url = {https://www.sciencedirect.com/science/article/pii/0550321386902877},
author = {L. Dixon and J. Harvey and C. Vafa and E. Witten},
}

@article{Candelas:1985en,
    author = "Candelas, P. and Horowitz, Gary T. and Strominger, Andrew and Witten, Edward",
    title = "{Vacuum configurations for superstrings}",
    reportNumber = "NSF-ITP-84-170",
    doi = "10.1016/0550-3213(85)90602-9",
    journal = "Nucl. Phys. B",
    volume = "258",
    pages = "46--74",
    year = "1985"
}

@article{Ellis:2005qa,
    author = "Ellis, John R. and Mavromatos, N. E. and Nanopoulos, Dimitri V. and Westmuckett, Michael",
    title = "{Liouville cosmology at zero and finite temperatures}",
    eprint = "gr-qc/0508105",
    archivePrefix = "arXiv",
    reportNumber = "CERN-PH-TH-2005-154, ACT-06-05, MIFP-05-19",
    doi = "10.1142/S0217751X06028990",
    journal = "Int. J. Mod. Phys. A",
    volume = "21",
    pages = "1379--1444",
    year = "2006"
}

@article{Nozawa:2005eu,
    author = "Nozawa, Masato and Maeda, Kei-ichi",
    title = "{Energy extraction from higher dimensional black holes and black rings}",
    eprint = "hep-th/0502166",
    archivePrefix = "arXiv",
    doi = "10.1103/PhysRevD.71.084028",
    journal = "Phys. Rev. D",
    volume = "71",
    pages = "084028",
    year = "2005"
}

@article{Banerjee:2010ye,
    author = "Banerjee, Rabin and Majhi, Bibhas Ranjan and Modak, Sujoy Kumar and Samanta, Saurav",
    title = "{Killing Symmetries and Smarr Formula for Black Holes in Arbitrary Dimensions}",
    eprint = "1007.5204",
    archivePrefix = "arXiv",
    primaryClass = "gr-qc",
    doi = "10.1103/PhysRevD.82.124002",
    journal = "Phys. Rev. D",
    volume = "82",
    pages = "124002",
    year = "2010"
}

@article{Modak:2010fn,
    author = "Modak, Sujoy Kumar and Samanta, Saurav",
    title = "{Effective Values of Komar Conserved Quantities and Their Applications}",
    eprint = "1006.3445",
    archivePrefix = "arXiv",
    primaryClass = "gr-qc",
    doi = "10.1007/s10773-011-1017-2",
    journal = "Int. J. Theor. Phys.",
    volume = "51",
    pages = "1416--1424",
    year = "2012"
}

@article{1972ApJ...178..347B,
       author = "Bardeen, James M. and Press, William H. and Teukolsky, Saul A.",
        title = "Rotating Black Holes: Locally Nonrotating Frames, Energy Extraction, and Scalar Synchrotron Radiation",
      journal = "apj",
         year = "1972",
        month = "dec",
       volume = "178",
        pages = "347-370",
          doi = "10.1086/151796",
       adsurl = "https://ui.adsabs.harvard.edu/abs/1972ApJ...178..347B"
}

@article{Maldacena:1997cg,
    author = "Maldacena, Juan Martin and Strominger, Andrew",
    title = "{Semiclassical decay of near extremal five-branes}",
    eprint = "hep-th/9710014",
    archivePrefix = "arXiv",
    doi = "10.1088/1126-6708/1997/12/008",
    journal = "JHEP",
    volume = "12",
    pages = "008",
    year = "1997"
}

@article{Aharony:1998ub,
    author = "Aharony, Ofer and Berkooz, Micha and Kutasov, David and Seiberg, Nathan",
    title = "{Linear dilatons, NS five-branes and holography}",
    eprint = "hep-th/9808149",
    archivePrefix = "arXiv",
    reportNumber = "EFI-98-39, RU-98-38, IASSNS-HEP-98-75",
    doi = "10.1088/1126-6708/1998/10/004",
    journal = "JHEP",
    volume = "10",
    pages = "004",
    year = "1998"
}

@article{Wald:1974kya,
    author = "Wald, Robert M.",
    title = "{Energy Limits on the Penrose Process}",
    doi = "10.1086/152959",
    journal = "Astrophys. J.",
    volume = "191",
    pages = "231",
    year = "1974"
}

@article{Christodoulou:1970wf,
    author = "Christodoulou, D.",
    title = "{Reversible and irreversible transforations in black hole physics}",
    doi = "10.1103/PhysRevLett.25.1596",
    journal = "Phys. Rev. Lett.",
    volume = "25",
    pages = "1596--1597",
    year = "1970"
}

@article{Caldarelli:1999xj,
    author = "Caldarelli, Marco M. and Cognola, Guido and Klemm, Dietmar",
    title = "{Thermodynamics of Kerr-Newman-AdS black holes and conformal field theories}",
    eprint = "hep-th/9908022",
    archivePrefix = "arXiv",
    reportNumber = "UTF-434",
    doi = "10.1088/0264-9381/17/2/310",
    journal = "Class. Quant. Grav.",
    volume = "17",
    pages = "399--420",
    year = "2000"
}

@article{Andrade:2018rcx,
    author = "Andrade, Tom{\'a}s and Emparan, Roberto and Licht, David",
    title = "{Charged rotating black holes in higher dimensions}",
    eprint = "1810.06993",
    archivePrefix = "arXiv",
    primaryClass = "hep-th",
    doi = "10.1007/JHEP02(2019)076",
    journal = "JHEP",
    volume = "02",
    pages = "076",
    year = "2019"
}

@article{Caldarelli:2010xz,
    author = "Caldarelli, Marco M. and Emparan, Roberto and Van Pol, Bert",
    title = "{Higher-dimensional Rotating Charged Black Holes}",
    eprint = "1012.4517",
    archivePrefix = "arXiv",
    primaryClass = "hep-th",
    reportNumber = "CPHT-RR111.1210, LPT-ORSAY-10-106",
    doi = "10.1007/JHEP04(2011)013",
    journal = "JHEP",
    volume = "04",
    pages = "013",
    year = "2011"
}

@article{Andrade:2018nsz,
    author = "Andrade, Tom\'as and Emparan, Roberto and Licht, David",
    title = "{Rotating black holes and black bars at large D}",
    eprint = "1807.01131",
    archivePrefix = "arXiv",
    primaryClass = "hep-th",
    doi = "10.1007/JHEP09(2018)107",
    journal = "JHEP",
    volume = "09",
    pages = "107",
    year = "2018"
}

@article{Penrose:1971uk,
    author = "Penrose, R. and Floyd, R. M.",
    title = "{Extraction of rotational energy from a black hole}",
    doi = "10.1038/physci229177a0",
    journal = "Nature",
    volume = "229",
    pages = "177--179",
    year = "1971"
}

@article{Myers:1986un,
    author = "Myers, Robert C. and Perry, M. J.",
    title = "{Black Holes in Higher Dimensional Space-Times}",
    reportNumber = "PRINT-86-0067 (PRINCETON)",
    doi = "10.1016/0003-4916(86)90186-7",
    journal = "Annals Phys.",
    volume = "172",
    pages = "304",
    year = "1986"
}

@inbook{Myers:2011yc,
    author = "Myers, Robert C.",
    editor = "Horowitz, Gary T.",
    title = "{Myers\textendash{}Perry black holes}",
    booktitle = "{Black holes in higher dimensions}",
    eprint = "1111.1903",
publisher="",
    archivePrefix = "arXiv",
    primaryClass = "gr-qc",
    pages = "101--133",
    year = "2012"
}

@article{Wald:1993nt,
    author = "Wald, Robert M.",
    title = "{Black hole entropy is the Noether charge}",
    eprint = "gr-qc/9307038",
    archivePrefix = "arXiv",
    reportNumber = "EFI-93-42",
    doi = "10.1103/PhysRevD.48.R3427",
    journal = "Phys. Rev. D",
    volume = "48",
    number = "8",
    pages = "R3427--R3431",
    year = "1993"
}

@article{schneiderbauer2020semi,
  title={Semi-Classical Black Hole Holography},
  author={Schneiderbauer, Lukas and others},
  year={2020},
  publisher={University of Iceland, School of Engineering and Natural Sciences, Faculty~?}
}

@article{Emparan:2013xia,
    author = "Emparan, Roberto and Grumiller, Daniel and Tanabe, Kentaro",
    title = "{Large-D gravity and low-D strings}",
    eprint = "1303.1995",
    archivePrefix = "arXiv",
    primaryClass = "hep-th",
    reportNumber = "TUW-13-04",
    doi = "10.1103/PhysRevLett.110.251102",
    journal = "Phys. Rev. Lett.",
    volume = "110",
    number = "25",
    pages = "251102",
    year = "2013"
}

@article{Soda:1993xc,
    author = "Soda, J.",
    title = "{Hierarchical dimensional reduction and gluing geometries}",
    doi = "10.1143/PTP.89.1303",
    journal = "Prog. Theor. Phys.",
    volume = "89",
    pages = "1303--1310",
    year = "1993"
}

@article{TEITELBOIM198341,
title = "Gravitation and hamiltonian structure in two spacetime dimensions",
journal = "Physics Letters B",
volume = "126",
number = "1",
pages = "41 - 45",
year = "1983",
issn = "0370-2693",
doi = "https://doi.org/10.1016/0370-2693(83)90012-6",
url = "http://www.sciencedirect.com/science/article/pii/0370269383900126",
author = "Claudio Teitelboim"
}

@article{JACKIW1985343,
title = "Lower dimensional gravity",
journal = "Nuclear Physics B",
volume = "252",
pages = "343 - 356",
year = "1985",
issn = "0550-3213",
doi = "https://doi.org/10.1016/0550-3213(85)90448-1",
url = "http://www.sciencedirect.com/science/article/pii/0550321385904481",
author = "Roman Jackiw"
}

@article{Russo:1992ax,
    author = "Russo, Jorge G. and Susskind, Leonard and Thorlacius, Larus",
    title = "{The Endpoint of Hawking radiation}",
    eprint = "hep-th/9206070",
    archivePrefix = "arXiv",
    reportNumber = "SU-ITP-92-17",
    doi = "10.1103/PhysRevD.46.3444",
    journal = "Phys. Rev. D",
    volume = "46",
    pages = "3444--3449",
    year = "1992"
}

@article{Schneiderbauer:2020isp,
    author = "Schneiderbauer, Lukas and Sybesma, Watse and Thorlacius, Lárus",
    title = "{Action Complexity for Semi-Classical Black Holes}",
    eprint = "2001.06453",
    archivePrefix = "arXiv",
    primaryClass = "hep-th",
    doi = "10.1007/JHEP07(2020)173",
    journal = "JHEP",
    volume = "07",
    pages = "173",
    year = "2020"
}

@article{Emparan:2020inr,
    author = "Emparan, Roberto and Herzog, Christopher P.",
    title = "{Large D limit of Einstein{\textquoteright}s equations}",
    eprint = "2003.11394",
    archivePrefix = "arXiv",
    primaryClass = "hep-th",
    doi = "10.1103/RevModPhys.92.045005",
    journal = "Rev. Mod. Phys.",
    volume = "92",
    number = "4",
    pages = "045005",
    year = "2020"
}

@article{Emparan:2014jca,
    author = "Emparan, Roberto and Suzuki, Ryotaku and Tanabe, Kentaro",
    title = "{Instability of rotating black holes: large D analysis}",
    eprint = "1402.6215",
    archivePrefix = "arXiv",
    primaryClass = "hep-th",
    doi = "10.1007/JHEP06(2014)106",
    journal = "JHEP",
    volume = "06",
    pages = "106",
    year = "2014"
}

@article{Sybesma:2022nby,
    author = "Sybesma, Watse",
    title = "{A zoo of deformed Jackiw-Teitelboim models near large dimensional black holes}",
    eprint = "2211.07927",
    archivePrefix = "arXiv",
    primaryClass = "hep-th",
    doi = "10.1007/JHEP01(2023)141",
    journal = "JHEP",
    volume = "01",
    pages = "141",
    year = "2023"
}

@article{Witten:1991yr,
    author = "Witten, Edward",
    title = "{On string theory and black holes}",
    reportNumber = "IASSNS-HEP-91-12",
    doi = "10.1103/PhysRevD.44.314",
    journal = "Phys. Rev. D",
    volume = "44",
    pages = "314--324",
    year = "1991"
}

@article{Mandal:1991tz,
    author = "Mandal, Gautam and Sengupta, Anirvan M. and Wadia, Spenta R.",
    title = "{Classical solutions of two-dimensional string theory}",
    reportNumber = "IASSNS-HEP-91-10",
    doi = "10.1142/S0217732391001822",
    journal = "Mod. Phys. Lett. A",
    volume = "6",
    pages = "1685--1692",
    year = "1991"
}

@article{Gautason:2020tmk,
    author = "Gautason, Friðrik Freyr and Schneiderbauer, Lukas and Sybesma, Watse and Thorlacius, Lárus",
    title = "{Page Curve for an Evaporating Black Hole}",
    eprint = "2004.00598",
    archivePrefix = "arXiv",
    primaryClass = "hep-th",
    doi = "10.1007/JHEP05(2020)091",
    journal = "JHEP",
    volume = "05",
    pages = "091",
    year = "2020"
}

@article{Callan:1992rs,
    author = "Callan, Curtis G., Jr. and Giddings, Steven B. and Harvey, Jeffrey A. and Strominger, Andrew",
    title = "{Evanescent black holes}",
    eprint = "hep-th/9111056",
    archivePrefix = "arXiv",
    reportNumber = "UCSB-TH-91-54, EFI-91-67, PUPT-1294",
    doi = "10.1103/PhysRevD.45.R1005",
    journal = "Phys. Rev. D",
    volume = "45",
    number = "4",
    pages = "1005",
    year = "1992"
}

@article{Nakarachinda:2022gsb,
    author = "Nakarachinda, Ratchaphat and Promsiri, Chatchai and Tannukij, Lunchakorn and Wongjun, Pitayuth",
    title = "{Thermodynamics of black holes with R{\'e}nyi entropy from classical gravity}",
    eprint = "2211.05989",
    archivePrefix = "arXiv",
    primaryClass = "gr-qc",
    doi = "10.1016/j.nuclphysb.2025.116796",
    journal = "Nucl. Phys. B",
    volume = "1011",
    pages = "116796",
    year = "2025"
}

@article{Shaymatov:2024fle,
    author = "Shaymatov, Sanjar",
    title = "{Efficiency of magnetic Penrose process in higher dimensional Myers-Perry black hole spacetimes}",
    eprint = "2402.02471",
    archivePrefix = "arXiv",
    primaryClass = "gr-qc",
    doi = "10.1103/PhysRevD.110.044042",
    journal = "Phys. Rev. D",
    volume = "110",
    number = "4",
    pages = "044042",
    year = "2024"
}

@article{Ferko:2021bym,
    author = "Ferko, Christian and Satishchandran, Gautam and Sethi, Savdeep",
    title = "{Gravitational memory and compact extra dimensions}",
    eprint = "2109.11599",
    archivePrefix = "arXiv",
    primaryClass = "gr-qc",
    reportNumber = "EFI{\textendash}20-5",
    doi = "10.1103/PhysRevD.105.024072",
    journal = "Phys. Rev. D",
    volume = "105",
    number = "2",
    pages = "024072",
    year = "2022"
}

\end{document}